\documentclass[structabstract,bibyear,longauth]{aa}
\usepackage{graphicx}
\begin{document}
   \title{The Lyman Continuum escape fraction of galaxies at $z=3.3$ in the
VUDS-LBC/COSMOS field\fnmsep\thanks{Based on data obtained with
the European Southern Observatory Very Large
Telescope, Paranal, Chile, under Large Program 185.A-0791 and on
observations made at the Large Binocular Telescope
(LBT) at Mt. Graham (Arizona, USA).}}

\author{A. Grazian\inst{1}
\and E. Giallongo\inst{1}
\and R. Gerbasi\inst{2,1}
\and F. Fiore\inst{1}
\and A. Fontana\inst{1}
\and O. Le F\`evre\inst{3}
\and L. Pentericci\inst{1}
\and E. Vanzella\inst{4}
\and G. Zamorani\inst{4}
\and P. Cassata\inst{5}
\and B. Garilli\inst{6}
\and V. Le Brun\inst{3}
\and D. Maccagni\inst{6}
\and L.A.M. Tasca\inst{3}
\and R. Thomas\inst{3}
\and E. Zucca\inst{4}
\and R. Amor\'in\inst{1}
\and S. Bardelli\inst{4}
\and L.P. Cassar\`a\inst{6}
\and M. Castellano\inst{1}
\and A. Cimatti\inst{7}
\and O. Cucciati\inst{7,4}
\and A. Durkalec\inst{3}
\and M. Giavalisco\inst{8}
\and N. P. Hathi\inst{3}
\and O. Ilbert\inst{3}
\and B.C. Lemaux\inst{3}
\and S. Paltani\inst{9}
\and B. Ribeiro\inst{3}
\and D. Schaerer\inst{10,11}
\and M. Scodeggio\inst{6}
\and V. Sommariva\inst{7,4}
\and M. Talia\inst{7}
\and L. Tresse\inst{3}
\and D. Vergani\inst{6,4}
\and A. Bonchi\inst{1}
\and K. Boutsia\inst{1}
\and P. Capak\inst{12}
\and S. Charlot\inst{13}
\and T. Contini\inst{11}
\and S. de la Torre\inst{3}
\and J. Dunlop\inst{14}
\and S. Fotopoulou\inst{9}
\and L. Guaita\inst{1}
\and A. Koekemoer\inst{15}
\and C. L\'opez-Sanjuan\inst{16}
\and Y. Mellier\inst{13}
\and E. Merlin\inst{1}
\and D. Paris\inst{1}
\and J. Pforr\inst{3}
\and S. Pilo\inst{1}
\and P. Santini\inst{1}
\and N. Scoville\inst{12}
\and Y. Taniguchi\inst{17}
\and P.W. Wang\inst{3}
          }

   \offprints{A. Grazian, \email{andrea.grazian@oa-roma.inaf.it}}

\institute{INAF--Osservatorio Astronomico di Roma, Via Frascati 33,
I-00040, Monte Porzio Catone, Italy
\and
Dipartimento di Fisica, Universit\'a di Roma La Sapienza,
P.le A. Moro 2, I-00185 Roma, Italy
\and
Aix Marseille Universit\'e, CNRS, LAM (Laboratoire d'Astrophysique
de Marseille) UMR 7326, 13388, Marseille, France
\and
INAF--Osservatorio Astronomico di Bologna, via Ranzani,1, I-40127,
Bologna, Italy
\and
Instituto de Fisica y Astronom\'ia, Facultad de Ciencias, Universidad de
Valpara\'iso, Av. Gran Breta\~{n}a 1111, Playa Ancha, Valpara\'iso, Chile
\and
INAF--IASF, via Bassini 15, I-20133, Milano, Italy
\and
University of Bologna, Department of Physics and Astronomy (DIFA), V.le Berti
Pichat, 6/2 - I-40127, Bologna, Italy
\and
Astronomy Department, University of Massachusetts, Amherst, MA 01003, USA
\and
Department of Astronomy, University of Geneva
ch. d'Ecogia 16, CH-1290 Versoix, Switzerland
\and
Geneva Observatory, University of Geneva, ch. des Maillettes 51, CH-1290
Versoix, Switzerland
\and
Institut de Recherche en Astrophysique et Plan\'etologie - IRAP, CNRS,
Universit\'e de Toulouse, UPS-OMP, 14, avenue E. Belin, F31400
Toulouse, France
\and
Department of Astronomy, California Institute of Technology, 1200 E. California 
Blvd., Pasadena, CA 91125, USA
\and
Institut d'Astrophysique de Paris, UMR7095 CNRS,
Universit\'e Pierre et Marie Curie, 98 bis Boulevard Arago, 75014
Paris, France
\and
SUPA, Institute for Astronomy, University of Edinburgh, Royal Observatory,
Edinburgh, EH9 3HJ, United Kingdom
\and
Space Telescope Science Institute, 3700 San Martin Drive,
Baltimore, MD 21218, USA
\and
Centro de Estudios de F\'isica del Cosmos de Arag\'on, Teruel, Spain
\and
Research Center for Space and Cosmic Evolution, Ehime University,
Bunkyo-cho 2-5, Matsuyama 790-8577, Japan
}

   \date{Received Month day, year; accepted Month day, year}

   \authorrunning{Grazian et al.}
   \titlerunning{The LyC escape fraction of galaxies at z=3.3
in VUDS-LBC/COSMOS}

% \abstract{}{}{}{}{} 
% 5 {} token are mandatory
 
  \abstract
  % context heading (optional)
{
The ionizing Lyman continuum flux escaping from high redshift galaxies
into the intergalactic medium is a fundamental quantity to
understand the physical processes involved in the reionization
epoch. However, from an observational point of view, direct detections
of HI ionizing photons at high redshifts are feasible for
galaxies mainly in the interval $z\sim 3-4$.
}
  % aims heading (mandatory)
{
We have investigated a sample of star-forming galaxies at $z\sim 3.3$
in order to search for possible detections of Lyman continuum ionizing
photons escaping from galaxy halos.
}
  % methods heading (mandatory)
{
UV deep imaging in the COSMOS field obtained with the prime focus
camera LBC at the LBT telescope was used together with a catalog of
spectroscopic redshifts obtained by the VIMOS Ultra Deep Survey (VUDS)
to build a sample of 45 galaxies at $z\sim 3.3$ with $L>0.5L^*$. We
obtained deep LBC images of galaxies with spectroscopic redshifts in
the interval $3.27<z<3.40$ both in the R and deep U bands (magnitude
limit $U\sim 29.7$ at S/N=1). At these redshifts the R band samples
the non-ionizing 1500 {\AA} rest frame luminosity and the U band
samples the rest-frame spectral region just short-ward of the Lyman
edge at 912 {\AA}. Their flux ratio is related to the ionizing escape
fraction after statistical removal of the absorption by the
intergalactic medium along the line of sight.
}
  % results heading (mandatory)
{
A sub-sample of 10 galaxies apparently shows escape fractions $>28$\%
but a detailed analysis of their properties reveals that, with the
exception of two marginal detections ($S/N\sim 2$) in the U band, all
the other eight galaxies are most likely contaminated by the UV flux
of low redshift interlopers located close (in angular position) to the
high-z targets. The average escape fraction derived from the stacking
of the cleaned sample was constrained to $f^{rel}_{esc}<$ 2\%. The
implied hydrogen photo-ionization rate is a factor two lower than
that needed to keep the intergalactic medium ionized at $z\sim 3$, as
observed in the Lyman-$\alpha$ forest of high-z QSO spectra or by the
proximity effect.
}
  % conclusions heading (optional), leave it empty if necessary 
{{
These results support a scenario where high redshift, relatively
bright ($L\ge 0.5 L^*$) star-forming galaxies alone are unable to sustain
the level of ionization observed in the cosmic intergalactic medium at
$z\sim 3$. Star-forming galaxies at higher redshift and at fainter
luminosities ($L<<L^*$) can be the major contributors to the reionization of the
Universe only if their physical properties are subject to rapid
changes from $z\sim 3$ to $z\sim 6-10$. Alternatively, ionizing
sources could be discovered looking for fainter sources among the AGN
population at high redshift.}
}

\keywords{Galaxies: distances and redshift - Galaxies: evolution -
Galaxies: high redshift - Galaxies: photometry}

   \maketitle
%
%________________________________________________________________

\section{Introduction}

The reionization epoch marks a fundamental event in the thermal
history of the Universe, probably located in the redshift interval
$z=6-10$. The lower limit is derived from observations of the Gunn-Peterson
effect in z=6 QSO spectra (\cite{fan06}) while the upper end
comes from measurements of the Thomson scattering in the CMB
polarization map by Wilkinson Microwave Anisotropy Probe (WMAP,
\cite{wmap}) and Planck (\cite{planck2015}).
The time extension of the reionization processes can also be constrained
by the so-called kinetic Sunyaev-Zel'dovich (kSZ) effect.
Combining data from the South Pole Telescope (SPT) with the WMAP
results, \cite{zahn12} provided a 95\% c.l. limit to the duration of
reionization of $\Delta z \le 4.4$, placing the end of reionization at
$z=7.2$. While we are able
to approximately locate the reionization epoch, we still do not have a
clear perception of the sources providing the bulk of the UV photons
required to reionize the Universe. Prime suspects have been searched so
far among high redshift star forming galaxies or AGNs.

It is well known that bright QSOs are efficient producers of HI ionizing
photons ($\lambda\le 912$ \AA), and they can ionize large bubbles of
hydrogen and helium even at distances up to several Mpcs
(\cite{cowie09,prochaska09,worseck14a,stevans14}). However, their volume density
is apparently too low at high redshifts ($z>2$) to provide the cosmic
photo-ionization rate required to keep the intergalactic medium (IGM)
ionized at $z\sim 6$ (\cite{fan06}). Recent searches for low
luminosity AGNs at high redshifts by means of deep X-ray imaging are
gradually changing this scenario
(\cite{fiore12,giallongo12,giallongo15}) leaving room for important
contributions to the reionization by this fainter AGN population.

High redshift star forming galaxies can also be important contributors
of ionizing photons at $z\ge 2$ (\cite{hm12,dc15}). An intense star
formation activity is observed till $z\sim 8-10$ (see for example
\cite{mcleod14}), and the luminosity function (LF) of galaxies at $z\ge 4$ is
gradually steepening with redshift (\cite{finkelstein14}). Young
stellar populations inside high redshift galaxies are producing a
large amount of UV photons, but only a fraction of them can reach the
surrounding IGM.
An efficient way of parametrizing the emission of HI ionizing photons
w.r.t. the non-ionizing ones is the definition of the escape fraction
(\cite{steidel01}).
The escape fraction parameter is a
measurement of the fraction of HI ionizing photons which are not
absorbed by the interstellar medium (ISM) and are thus free to ionize
the neutral hydrogen in the IGM.
The extensive definition of the escape fraction will be provided in Sect. 3.

The bright star forming galaxies at low redshifts ($z\le 2$) in
general are not able to ionize their neighborhoods since their escape
fractions are only few percents or less
(\cite{grimes09,cowie09,siana10,bridge10,leitet13}). At high
redshifts the current situation is not so clear. The escape fractions
from $z\ge 3$ star-forming galaxies measured by different groups in
the last 15 years led to different results indicating that some
systematics are still present in the analysis and in the galaxy
samples used.

\cite{steidel01} claimed the detection of significant emission
($f^{rel}_{esc}\sim 50$\%) of Lyman Continuum (LyC) photons from the
composite spectrum of
29 Lyman break galaxies (LBG) at $z\sim 3$. Two LBGs from the
Steidel high-redshift sample were spectroscopically observed at
intermediate resolution by \cite{giallongo02} with very deep
integration. They derived only an upper limit to the relative escape
fraction of $f^{rel}_{esc}\le 16\%$. Further spectroscopic work by Steidel's
team found only two emitting galaxies at $z\sim 3$, out of 14
analyzed, with significant escape fraction ($f^{rel}_{esc}\ge 50-100\%$). For
the remaining population only upper limits were derived resulting in
an average $f^{rel}_{esc}\sim 14\%$ for the whole sample (\cite{shapley06}).
However, the two emitting galaxies were not confirmed as LyC emitters
by \cite{iwata09} and \cite{nestor11}. The original result of
\cite{shapley06} should thus be
converted into a strict upper limit of $f^{rel}_{esc}\le 2\%$ at $z\sim 3$
(\cite{cooke14}).

With the aid of narrow band imaging at $\lambda\sim 3590$ \AA,
\cite{iwata09} and \cite{nestor11} studied the same area of
\cite{shapley06}, namely the SSA22 field. They found a significant
population of LyC emitters at $z\ge 3$, with escape fractions in
the range $f^{rel}_{esc}\sim 4-100\%$ (some Lyman-$\alpha$ emitters with
$f^{rel}_{esc}\ge 100\%$), providing the first indication that the
escape fraction of galaxies started to increase at $z\ge 3$ (or that
their sample may be contaminated by foreground objects, as we will discuss
in the following). Recently, \cite{mostardi13} studied a new sky
region around an overdensity of sources at $z\sim 2.8$, reaching
slightly similar conclusions ($f^{rel}_{esc}\sim 5-49\%$).

Results from other teams based on very deep narrow or broad band UV
imaging gave only stringent upper limits of $f^{rel}_{esc}\le 5-10\%$ to the
escape fraction of star forming galaxies at $z=3-4$
(\cite{vanzella10b,boutsia11}). The main reason for the discrepant
results is due to the details of the photometric analysis and to the
different morphological analysis of the candidate Lyman continuum emitters.

The most convincing explanation for this discrepancy has been proposed
by \cite{vanzella10a} (hereafter V10a) on the basis of possible UV flux
contamination in the Nestor's and Mostardi's LyC emitting
sample by low redshift interlopers which are near the position
of the high redshift galaxies. As estimated by V10a, the chance for
contamination increases with large seeing and deep UV images.
Considering a sample of 101 galaxies at $3.4\le z\le 4.4$ in the
GOODS-South field they found only one convincing detection after
removing probable contaminants identified on the basis of accurate
multicolor HST photometry at high angular resolution.

At the same time \cite{boutsia11} studied a small sample of 11
galaxies at $z\sim 3.3$ in the Q0933, Q1623, and COSMOS fields with
very deep imaging in the U and R bands obtained at the prime focus camera
LBC of LBT. The non detection of all the galaxies resulted in a
strict upper limit to the escape fraction of $f^{rel}_{esc}<5$\%, but the
result was affected by low number statistics and by the uncertainty
due to the stochasticity of the IGM absorption. Their conclusions were
that, at present, a class of ionizing sources responsible for cosmic
reionization has not been clearly identified and a step forward to the
escape fraction studies is required.

Recently, \cite{siana15} re-analyzed with the HST instrument WFC3-UVIS
5 galaxies that were identified by \cite{nestor11} as LyC candidates at
$z\sim 3.1$. Thanks to deep F336W images and NIR Keck spectra
of these galaxies, \cite{siana15} concluded that the supposed LyC emission
is in fact provided by lower redshift interlopers. The new upper limit by
these observations is $f^{rel}_{esc}\sim 7-9\%$, consistent with the limits
by \cite{vanzella10b} and \cite{boutsia11}. In addition,
\cite{mostardi15} reported an HST follow up analysis of 16 candidate
LyC emitters found by \cite{mostardi13}. They obtained deep $U_{336}$,
$V_{606}$, $J_{125}$, and $H_{160}$ imaging with WFC3 and ACS for these galaxies:
thanks to the high spatial resolution of HST, they showed that 15 emitters
can be explained with contamination of low-z objects, wrong spectroscopic
redshifts or uncertain photometric redshifts. Only one galaxy shows a robust
LyC detection, according to \cite{mostardi15}. On the basis of these results,
they concluded that galaxies at $z\sim 3$ are providing the same contribution
of QSOs to the HI photo-ionizing background, although with large uncertainties.

Increasing the number of accurate LyC measures is not a trivial task
since it requires either very deep spectroscopy or very deep UV
imaging of a large sample of galaxies with known spectroscopic
redshifts in the range $z=3-4$ where the IGM still allows the
transmission of some ionizing flux along the line of sight
(\cite{inoue14}). Indeed at higher redshifts it is difficult to
disentangle the absorption due to the galaxy ISM from that due to the
surrounding IGM along the line of sight (LoS).

In this work we exploit the availability of very deep U-band photometry
obtained with the LBC prime focus imager at LBT, coupled with a large number of
galaxies with spectroscopic redshifts in a relatively narrow
range around $z\simeq 3.3$ that allows the UV band to sample the
spectral region just short-ward of 912 {\AA} rest. The spectroscopic
sample is taken from the VUDS survey (\cite{vuds}) of high redshift
galaxies ($2<z<\sim 6$) in the COSMOS field where deep HST imaging
is also available and allows to look for possible contamination by low
redshift interlopers.

This paper is organized as follows. In Sect. 2 we present the dataset,
in Sect. 3 we describe the method adopted, in Sect. 4 and 5 we show
the results for individual objects and for the overall sample as a
whole, in Sect. 6 we discuss our results
and in Sect. 7 we provide a summary. Throughout the paper
we adopt the $\Lambda$-CDM concordance cosmological model ($H_0 = 70~
km/s/Mpc$, $\Omega_M=0.3$ and $\Omega_\Lambda=0.7$). All magnitudes
are in the AB system.

%__________________________________________________________________

\section{Data}

The COSMOS field (\cite{scoville}) has been selected as an optimal
area for the measurement of the escape fraction of galaxies at $z\sim
3$, due to its combination of deep multiwavelength imaging by LBC
coupled with a large number of spectroscopic surveys over a broad
redshift interval $0<z<7$ obtained by extensive observational
campaigns (\cite{zcosmos,vuds}).

In the present analysis we have also included the galaxies already
studied by \cite{boutsia11} in the Q0933 and Q1623 fields.

\subsection{The photometric dataset}

The imaging data set used in this paper consists of deep LBC
observations of the COSMOS field in the U and R filters. These deep
observations are part of a more ambitious program, the LBC/CANDELS
(\cite{grazLBC}) survey, with the goal of providing complementary deep
UV coverage with LBC to the northern fields (COSMOS, GOODS-North, EGS)
of the CANDELS survey (\cite{grogin11,koekemoer11}).
The LBC imaging dataset of the COSMOS field has been described in
detail in \cite{boutsia11} and \cite{boutsia14}.

The LBC instrument (\cite{giallongo08}) has two wide-field imagers,
one optimized for the UV wavelengths and the other for the optical-red
bands. This imager has observed part ($\sim 0.20$ sq. deg.) of the
COSMOS (\cite{scoville}) field in UGRIZ bands to search for LBGs at
$2.7<z<3.4$ (\cite{boutsia14}). These data, obtained with one of the
most efficient ground-based large field UV imager at an 8 meter class
telescope, the LBT, have allowed us to derive a stringent limit of
$\le 5\%$ to the escape fraction of 11 star-forming galaxies at $z\sim 3.3$
(\cite{boutsia11}), using the same technique that will be adopted in
this work.

The U-band observations of the COSMOS field by LBC are constituted by three
partially overlapping pointings of $\sim 6$ hours each, for a total of 11
hours of exposure on the deepest regions. The relative good seeing ($\sim
1.0$ arcsec) and the high sensitivity of the LBC camera in the UV allow
us to reach a depth of 28.7 AB magnitudes (at S/N=1) over an area
corresponding to 700 sq. arcmin., with slightly deeper limits ($\sim 29.1$)
on the regions covered twice ($\sim 450$ sq. arcmin).

The R band dataset has an average seeing of 0.9 arcsec and in 3.3 hours
of exposure time it reaches, for two third of the area covered by deep
U band imaging, a depth of R=25.7 AB magnitude at S/N=10. The third
pointing is shallower (R=23.8), since the exposure time was shorter
(450 seconds). In this sub-region, we select only galaxies down to
$R\le 24.0$ to work at a good S/N ratio in the detection band, which
will be used to measure the non ionizing continuum of the sources.

The standard LBC pipeline has been used to reduce the imaging dataset,
as described in detail in \cite{giallongo08,boutsia11} and
\cite{boutsia14}. In particular, after correcting the images for bias
subtraction, we have applied standard flat-fielding using sky flats
obtained during twilight. The sky background was estimated using
SExtractor (\cite{sex}) with a mesh of 64 by 64 pixels and
adopting a median of 3 by 3 meshes. The astrometric corrections
were computed using the software ``AstromC'' described by \cite{radovich04}.
After resampling all the individual images in a common
reference frame, we co-added the various images to create deep mosaics
for each field.

For each LBC image, we have derived a careful estimate of the absolute
rms map taking into account the Poisson statistics and the
instrumental gain for each individual chip, as described in
\cite{boutsia11}. We have then propagated the rms map over the whole
data reduction chain, obtaining as a final product of the pipeline an
absolute variance map associated with each stacked scientific frame.

\subsection{Spectroscopic redshifts}

The VUDS spectroscopic survey (\cite{vuds}) is an ESO Large Program
(P.I. Le F\`evre)
aiming at obtaining $\sim$10,000 spectroscopic redshifts with
the VIMOS instrument to study early phases of galaxy evolution in the
redshift range $2<z<6$ and beyond. This
ambitious program has been conducted over 3 fields (VVDS-02h, COSMOS,
ECDFS) and the targets were selected with different criteria
(inclusive combination of U-dropout, B-dropout, and photometric
redshift selections), tailored to maximize the efficiency of finding
galaxies at $3<z<5$, thus being complementary to the zCOSMOS survey
(\cite{zcosmos}). The spectra are the
result of deep integration (14 hours per target) covering the
wavelength range $3650\le \lambda\le 9350$ \AA.
Details on the observational strategy, candidate selection, data reduction,
quality flag assignment
and first results of the VUDS survey can be found in \cite{vuds}.
The data acquisition, reduction and redshift extraction procedure of the
whole VUDS survey have been completed recently.

We focus on a very small redshift range ($3.27<z<3.40$) in order
to minimize the effect of the strong IGM opacity evolution with
redshift (\cite{inoue14,worseck14a}) and allowing a
direct measurement of the relative ionizing escape fraction from star
forming galaxies at $z=3.3$. As explained also in \cite{boutsia11},
this narrow redshift range allows us to use the LBC U-band filter
(SDT-Uspec) to sample the rest frame wavelengths short-ward of 912 \AA,
but not too far from it, namely at $\ge 880$ {\AA} rest frame.
An extension of the redshift range at $z>3.40$ is possible, but
it requires detail simulations
of the variance due to the stochasticity of intervening damped
Lyman-$\alpha$ systems or Lyman Limit systems, as shown in \cite{vanzella10b}.
Since it goes beyond the scope of this paper,
we limit the analysis to $z<3.40$.

We started from the sample of 73 galaxies with highly reliable
(flag=3, 4, 23, 24) spectroscopic redshifts within the interval
$3.27<z<3.40$ in the COSMOS area covered by the VUDS spectroscopic
survey. We do not consider within our sample the AGNs from the optical
classification of the VUDS spectra (flag=11, 12, 13, 14, 19 or 211,
212, 213, 214, 219 due mainly to the presence of a strong CIV emission
line). We excluded also the spectroscopic redshifts based only on a
single emission line (flag=9 or 29). Of the pure (non AGN) galaxies,
only 35 are covered by the LBC images in the U and R bands, which do
not cover the whole COSMOS field, as mentioned above. One of these
sources has been excluded since it falls on an horizontal stripe of
bad pixels caused by a saturated star in the LBC images. Eleven
additional galaxies were already discussed by \cite{boutsia11},
resulting in a total sample of 45 star forming galaxies at $z\sim 3.3$
with secure spectroscopic redshifts and deep U and R band images by
LBC. Table \ref{table:gal} summarizes the properties of the individual
galaxies which are used in this paper to study the escape fraction at
$z=3.3$.

%__________________________________________________________________

\section{The adopted method to derive the LyC escape fraction of star-forming
galaxies in the VUDS-LBC/COSMOS field}

In this paper we exploit the deep UV imaging by LBC in the
VUDS/COSMOS field to derive a meaningful measurement of the ionizing
radiation of star forming galaxies at $z\sim 3.3$.

The LyC escape fraction of high-z galaxies can be evaluated by
adopting very deep UV imaging data sampling the $\lambda\le 912$ {\AA}
rest frame (Lyman continuum) region. Since the intrinsic spectral
energy distribution (SED) of a galaxy is not known a priori,
especially in the rest-frame far UV where significant dust reddening
could be present, a relative escape fraction is usually derived from
the observations (\cite{steidel01}). It is defined as the fraction of
escaping Lyman continuum photons divided by the fraction of escaping
photons at 1500 {\AA} rest frame (\cite{siana07}).

The SED of a high-redshift source at $\lambda\le 912$ {\AA} rest frame
is also affected by the photoelectric absorption by the IGM
(\cite{moller90,madau95}) and it is mainly due to Lyman Limit Systems (LLS)
with column densities of neutral hydrogen
between $log(N_{HI})\sim 17$ and 20. To correct
for the mean effect of the IGM absorption we adopt the recent
estimates by \cite{worseck14a}. Convolving the IGM
absorption at $z=3.3$ with the U-band filter of LBC and adopting the
same procedure outlined in \cite{boutsia11} we obtain an IGM mean
transmission $<exp(-\tau_{900}^{IGM})>=0.28$ at $z\sim 3.3$, similar
to that found adopting the IGM corrections by \cite{prochaska09}.

It is worth noting that we are not considering here the variance of the IGM
absorption in different lines of sight due to a patchy pattern of
LLS. Our relatively large number of galaxies allows us to deal,
from the statistical point of view, with the possibility that few
lines of sight are completely absorbed by intervening optically thick
systems at $z\le 3.3$. In the discussion section we will quantify this
probability.

Following \cite{steidel01} and \cite{siana07},
we estimate the relative escape fraction
using the following equation:

\begin{equation}
f^{rel}_{esc}=\frac{(L_{1500}/L_{900})_{int}}{(F_R/F_U)_{obs}}exp(\tau_{900}^{IGM}) \,
\end{equation}

which is equivalent to:

\begin{equation}
f^{rel}_{esc}=\exp(-\tau_{HI})\times 10^{-0.4(A_{900}-A_{1500})}=f^{abs}_{esc}\times
10^{0.4A_{1500}}
\end{equation}

where $(L_{1500}/L_{900})_{int}$ is the ratio of the intrinsic luminosities at
1500 and 900 {\AA} rest frame, $(F_R/F_U)_{obs}$ is the ratio of the
observed fluxes in the R and U band, $\exp(-\tau_{HI})$ indicates the
ISM extinction,
$\exp(-\tau_{HI})\times 10^{-0.4 A_{900}}$ is the absolute escape fraction
$f^{abs}_{esc}$, $A_{1500}$ and $A_{900}$
are the dust absorption coefficients at 1500 and 900 \AA, respectively.
According to \cite{Calzetti2000}, $A_{1500}=10.33 E(B-V)$ and
$A_{900}=18.02 E(B-V)$, thus giving $f^{rel}_{esc}=2.59 f^{abs}_{esc}$ for
$E(B-V)=0.1$. According to Eq. 2, this implies that $f^{rel}_{esc}$ is always
less than 100\%.

We adopt an intrinsic ratio of $(L_{1500}/L_{900})_{int}=3$
for our star-forming galaxies, taking into account the absorption
produced by the stellar atmospheres, as adopted by \cite{steidel01}
and \cite{boutsia11} on the basis of predictions by the spectral
synthesis stellar models. The intrinsic ratio
$(L_{1500}/L_{900})_{int}$ depends on the physical properties of the
galaxies, mainly the mean stellar ages, metallicities, IMFs, and
SFHs. For typical star forming galaxies at $z\sim 3.3$
the intrinsic ratio $(L_{1500}/L_{900})_{int}$ varies between 1.7 and 7.1 for
age between 1 Myr and 0.2 Gyr, adopting the BC03
library (\cite {bc03}).
In the following sections we will discuss the implications of
our choice, and we will show that our final results on the ionization
emissivity are robust against
the adopted value of 3 for the intrinsic ratio
$(L_{1500}/L_{900})_{int}$. To derive the observed fluxes at 900 and
1500 {\AA} rest frame at $z\sim 3.3$ we rely on the U and R band flux
ratio measured by LBC, $(F_R/F_U)_{obs}$.

In our previous paper (\cite{boutsia11}), we have estimated fluxes
using aperture photometry matched to the seeing of the U and R
bands. Here we explore a slightly different approach: if LyC photons
are emitted by a particular region of a galaxy characterized by low
ISM absorption, then we must observe also non-ionizing emission
emerging from the {\em same} spatial region (see section 4.1.1 of
\cite{vanzella12}, hereafter V12). As a consequence, the emission in the U band
(sampling $\sim 900$ {\AA} at $z\sim 3.3$) must be always associated with
an emission in the R band ($\sim 1500$ {\AA} at $z\sim 3.3$). The
opposite would not be true, however, since non-ionizing radiation can
be detected in regions of a galaxy where the ISM or dust absorption
are elevated, with a negligible emission of LyC photons. The observed
profile of a galaxy in the R band can thus be used to measure the
ionizing radiations in the U band only in the physical regions where
such photons are expected, limiting the contamination by foreground
galaxies lying by chance on a given line of sight of a star forming
galaxy at $z\sim 3.3$.
We thus adopt the ConvPhot software (\cite{desantis07}) to
derive less contaminated photometry in the U band, using the R-band image,
with its much higher S/N ratio, as a prior to reconstruct the profile
of galaxies in the U band. Although the computed magnitude is not
significantly different from that derived by aperture photometry, the
photometric uncertainties are smaller in the ConvPhot procedures since
the flux errors in the U band are, at least at a first approximation, equal
to the result of the sum over
its $rms^2$ weighing for the profile of the object in the R band.
Comparing the objects in common with \cite{boutsia11}, we
find that this new method allows a gain by a factor of $\sim$1.7 in
S/N ratio in the U band photometry, and thus a similar improvement on
the determination of the uncertainties on the LyC escape fraction.
The resulting limiting magnitude in the LBC U band image is thus $\sim
29.7$ AB at S/N=1 for the central overlapping region, and slightly
shallower ($\sim 29.3$) in the flanking areas covered with only one
LBC pointing.

Following V12, we define ``local'' escape fraction as the parameter
measuring the LyC emission w.r.t. non ionizing radiation, measured in
the same physical region where both the 900 and 1500 {\AA} rest frame
wavelengths are emitted.  As a sanity check for the consistency of the
following results, the ``local'' escape fraction cannot exceed 100\%
for reasonable values of the IGM transmission and of the intrinsic
ratio $(L_{1500}/L_{900})_{int}$. As in V12, a ``local'' escape fraction
of 500-1000\% can be a clear indication of contamination in the LyC band
from a foreground interloper.

%__________________________________________________________________

\section{The relative escape fraction of individual galaxies at $z\sim 3.3$}

Using the method outlined in the previous section, we computed the relative
escape fraction of the 45 individual galaxies at $3.27\le z_{spec}\le 3.40$
of the VUDS-LBC/COSMOS sample. In case of non detection in the U band
(i.e. $S/N<1$) we provide the upper limit of the relative escape fraction at
1$\sigma$. This is a fair limit since we are forcing the measurement of LyC
flux in a region where we see radiation at $\lambda\sim 1500$
{\AA} rest frame (non ionizing) from the clear detection in the R band.
Table \ref{table:gal} summarizes the
properties of these objects. As we will discuss in section 6.2, the patchy
nature of the IGM makes the determination of the escape fraction for
individual galaxies slightly uncertain. Our main results has thus been derived
in section 5.1 by studying the properties of the whole galaxy population
at $z\sim 3$ through image stacking.

%__________________________________________________Two columns table
\begin{table*}
\caption{The galaxies in the VUDS-LBC/COSMOS field used to derive the LyC
relative escape fraction}
\label{table:gal}
\centering
\begin{tabular}{r | c c c c c c c c c}
\hline
\hline
$ID_{LBC}$ & $Field$ & $RAD$ & $Dec$ & $z_{spec}$ & $Rmag$ & $Umag$ & $M_{1500}$ & $f^{rel}_{esc}$ & $Cont$ \\
   &     & (deg)    & (deg)      &     &       (AB) & (AB) & &      &           \\
\hline
13903 & COSMOS & 150.414993286 & +2.159081936 & 3.290 & 22.59 & $\ge$ 28.69 & -23.11 & $\le$ 0.039 & No \\
74113 & COSMOS & 149.886260986 & +2.276002169 & 3.330 & 23.56 & $\ge$ 29.50 & -22.14 & $\le$ 0.045 & No \\
56982 & COSMOS & 149.847152710 & +2.373027802 & 3.356 & 23.65 & $\ge$ 29.46 & -22.05 & $\le$ 0.051 & No \\
72509 & COSMOS & 149.994003296 & +2.285038948 & 3.374 & 23.84 & $\ge$ 29.38 & -21.86 & $\le$ 0.065 & No \\
16700 & COSMOS & 149.940277100 & +2.082487583 & 3.354 & 24.44 & $\ge$ 29.83 & -21.26 & $\le$ 0.075 & No \\
61293 & COSMOS & 149.870300293 & +2.349019766 & 3.292 & 23.90 & $\ge$ 29.13 & -21.80 & $\le$ 0.087 & No \\
21413 & COSMOS & 149.974533081 & +2.111910582 & 3.319 & 24.50 & $\ge$ 29.68 & -21.20 & $\le$ 0.091 & No \\
31070 & COSMOS & 149.912933350 & +2.236840248 & 3.333 & 24.74 & $\ge$ 29.86 & -20.96 & $\le$ 0.096 & No \\
64371 & COSMOS & 150.051940918 & +2.332614422 & 3.370 & 24.19 & $\ge$ 29.26 & -21.51 & $\le$ 0.100 & No \\
49197 & COSMOS & 149.978881836 & +2.135632515 & 3.395 & 24.81 & $\ge$ 29.80 & -20.89 & $\le$ 0.108 & No \\
58808 & COSMOS & 149.962997437 & +2.363846302 & 3.359 & 24.63 & $\ge$ 29.58 & -21.07 & $\le$ 0.112 & No \\
68731 & COSMOS & 150.138900757 & +2.307063818 & 3.288 & 24.47 & $\ge$ 29.41 & -21.23 & $\le$ 0.113 & No \\
32388 & COSMOS & 149.778930664 & +2.229439735 & 3.300 & 24.63 & $\ge$ 29.53 & -21.07 & $\le$ 0.118 & No \\
16446 & COSMOS & 149.961273193 & +2.081201315 & 3.270 & 24.96 & $\ge$ 29.86 & -20.74 & $\le$ 0.118 & No \\
69814 & COSMOS & 149.834243774 & +2.300836325 & 3.366 & 24.56 & $\ge$ 29.45 & -21.14 & $\le$ 0.119 & No \\
 3400 & Q0933  & 143.354507446 & +28.806911469 & 3.270 & 24.88 & $\ge$ 29.76 & -20.82 & $\le$ 0.120 & No \\
 1723 & COSMOS & 150.447082520 & +2.347564697 & 3.300 & 23.63 & $\ge$ 28.43 & -22.07 & $\le$ 0.129 & No \\
56248 & COSMOS & 149.917510986 & +2.378670692 & 3.288 & 24.71 & $\ge$ 29.50 & -20.99 & $\le$ 0.130 & No \\
63275 & COSMOS & 149.983078003 & +2.338610888 & 3.343 & 24.96 & $\ge$ 29.72 & -20.74 & $\le$ 0.134 & No \\
72124 & COSMOS & 149.883422852 & +2.286791325 & 3.352 & 24.71 & $\ge$ 29.46 & -20.99 & $\le$ 0.135 & No \\
51227 & COSMOS & 149.892135620 & +2.414754868 & 3.280 & 24.45 & $\ge$ 29.18 & -21.25 & $\le$ 0.137 & No \\
50989 & COSMOS & 149.834274292 & +2.416670561 & 3.310 & 24.61 & $\ge$ 29.32 & -21.09 & $\le$ 0.140 & No \\
51473 & COSMOS & 150.065383911 & +2.411855936 & 3.355 & 25.29 & $\ge$ 29.90 & -20.41 & $\le$ 0.153 & No \\
 8556 & Q0933  & 143.382461548 & +28.753007889 & 3.330 & 24.91 & $\ge$ 29.51 & -20.79 & $\le$ 0.155 & No \\
17175 & Q1623  & 246.469177246 & +26.892374039 & 3.340 & 24.69 & $\ge$ 29.29 & -21.01 & $\le$ 0.155 & No \\
47539 & COSMOS & 149.888641357 & +2.144912004 & 3.311 & 25.02 & $\ge$ 29.59 & -20.68 & $\le$ 0.159 & No \\
30576 & COSMOS & 150.141159058 & +2.239602089 & 3.336 & 25.26 & $\ge$ 29.82 & -20.44 & $\le$ 0.160 & No \\
10235 & COSMOS & 150.106704712 & +2.044517517 & 3.328 & 25.14 & $\ge$ 29.70 & -20.56 & $\le$ 0.161 & No \\
70238 & COSMOS & 149.853439331 & +2.298267603 & 3.292 & 24.58 & $\ge$ 29.13 & -21.12 & $\le$ 0.162 & No \\
42374 & COSMOS & 150.137237549 & +2.174738884 & 3.294 & 25.44 & $\ge$ 29.87 & -20.26 & $\le$ 0.181 & No \\
19739 & COSMOS & 149.934432983 & +2.102468014 & 3.316 & 25.54 & $\ge$ 29.96 & -20.16 & $\le$ 0.183 & No \\
12646 & Q0933  & 143.328750610 & +28.719076157 & 3.330 & 25.13 & $\ge$ 29.50 & -20.57 & $\le$ 0.192 & No \\
10849 & Q0933  & 143.360122681 & +28.734085083 & 3.350 & 25.59 & $\ge$ 29.92 & -20.11 & $\le$ 0.198 & No \\
65540 & COSMOS & 149.848281860 & +2.325203896 & 3.320 & 25.50 & $\ge$ 29.58 & -20.20 & $\le$ 0.250 & No \\
 4890 & COSMOS & 149.824325562 & +2.009310722 & 3.301 & 25.24 & $\ge$ 29.24 & -20.46 & $\le$ 0.268 & No \\
\hline
19521 & COSMOS & 150.330398560 & +2.072281122 & 3.303 & 22.55 & 26.50 & -23.15 & $0.281^{(a)}$ & Yes \\
63327 & COSMOS & 149.843582153 & +2.340100288 & 3.329 & 25.02 & 28.47 & -20.68 & $0.446^{(b)}$ & No\\
25155 & COSMOS & 149.857284546 & +2.270117283 & 3.364 & 25.24 & 28.61 & -20.46 & $0.479^{(c)}$ & No \\
40998 & COSMOS & 149.952316284 & +2.182848215 & 3.321 & 25.19 & 28.36 & -20.51 & $0.577^{(d)}$ & Yes \\
48734 & COSMOS & 149.815368652 & +2.138106108 & 3.384 & 24.99 & 27.99 & -20.71 & $0.674^{(e)}$ & Yes \\
47357 & COSMOS & 150.096496582 & +2.145905495 & 3.292 & 24.91 & 27.78 & -20.79 & $0.760^{(f)}$ & Yes \\
\hline
52336 & COSMOS & 149.980270386 & +2.405712605 & 3.371 & 24.85 & 27.06 & -20.85 & 1.400 & Yes \\
55594 & COSMOS & 149.939086914 & +2.388362885 & 3.334 & 24.17 & 26.06 & -21.53 & 1.873 & Yes \\
16669 & COSMOS & 149.994964600 & +2.080330372 & 3.367 & 24.39 & 26.06 & -21.31 & 2.295 & Yes \\
72687 & COSMOS & 149.877334595 & +2.283946991 & 3.328 & 24.14 & 25.55 & -21.56 & 2.920 & Yes \\
\hline
\hline
\end{tabular}
\\
The galaxies have been sorted in ascending order of relative escape
fraction. The upper part of the table summarizes the non detections
(i.e. $S/N<1$) in the LBC U-band, for which an upper limit to the
relative escape fraction has been provided. The central sub-sample
includes galaxies with apparent Lyman Continuum detection, while the
last four galaxies have escape fraction larger than 1.0 (i.e. $100\%$)
and are thus considered as unreliable and probably due to foreground
contamination, as discussed in the main text. The last column,
labeled $Cont$, indicates whether a galaxy is possibly contaminated by
foreground emission. Notes on individual objects: (a) local escape
fraction 1.75, calculated on the LyC emission region only; (b,c) they
can be plausible LyC emitters; (d) local escape fraction 4.51; (e)
local escape fraction 6.40; (f) local escape fraction 5.13.
\end{table*}

Table \ref{table:gal} shows that there are 10 detections in the U band within 
the initial sample of 45 star forming galaxies at $z\sim 3.3$ in the VUDS 
area covered by LBC observations (complemented with early results from 
\cite{boutsia11}). At face value, this result suggests possible 
LyC emission for this sub-sample.

\subsection{Galaxies with possible LyC detection}

We discuss here the properties of the ten interesting objects at $z\sim
3.3$ which are detected in the U band of LBC showing possible ionizing
emission. In the next sub-sections we check the reliability of the LyC
sources based on the deep HST F814W (hereafter indicated as I814W) band
and the LBC U and R bands.

\subsubsection{Galaxies with $f^{rel}_{esc}>100\%$}

In Table \ref{table:gal} there are four galaxies with a non-physical
escape fraction ranging between 140 and 292\% (last four objects at
the end of the table).

\begin{figure*}
\includegraphics[width=18cm,angle=0]{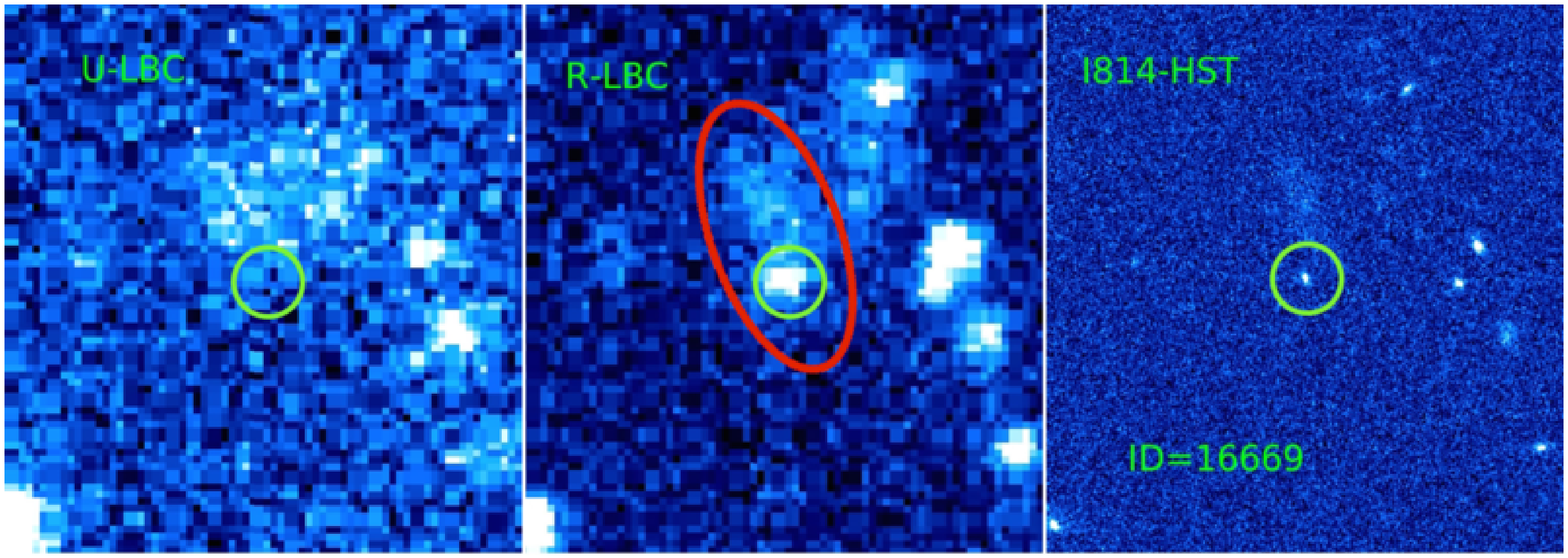}
\caption{The cutouts of source ID=16669 in the U band (left) and in the
R band (center) by the LBC instrument. The cutout on the right shows the
same sky area observed by HST in the I814W filter. The position of the
galaxy with spectroscopic redshift is indicated by the green circle.
A diffuse emission
at the north of the $z\sim 3.3$ galaxy is detected in all the three bands.
The red ellipse in the central panel indicates the area covered by source
ID=16669 in the LBC R band. This region has been used to compute the ``global''
escape fraction provided in the last column of Table \ref{table:gal}.
The size of each image is 20 $arcsec$ by 20 $arcsec$.
}
\label{obj16669ima}
\end{figure*}

Fig.\ref{obj16669ima} shows as an example the cutouts in the LBC U and R bands,
and the HST image in the I814W filter of the galaxy with ID=16669 in
Table \ref{table:gal}. The spectroscopic redshift from VUDS is
associated with a compact galaxy (indicated by the green circle in the HST
image on the right). The compact source falls near (distance=2.1 arcsec) a
bright and extended object which is clearly detected in the U band, while
at the position of the $z\sim 3.3$ galaxy no flux is emitted in the U band.
When computing the escape fraction with our method (i.e. detecting the
galaxy in the LBC R-band and measuring the LyC flux in the LBC U-band on the
same area, shown by the red ellipse in the central panel of
Fig.\ref{obj16669ima}), the two galaxies are
merged in a single detection in the R band, and thus our procedure gives
a significant LyC detection from the U band.
Translating
the observed U and R fluxes into an escape fraction, it turns out that this
object has $f^{rel}_{esc}=229.5\%$, more than twice the maximum value allowed
for a typical star forming galaxy. If we restrict
the investigated area to the region
with a significant U-band emission (the area of the red ellipse in
Fig.\ref{obj16669ima} outside the green circle; this barely extended object is
also visible in the HST image), and we compute the flux ratio between
the U and R band,
we end up with an escape fraction of 519.4\%. The uncertainties in the U and
R magnitudes are of the order of $\sim 0.05$, indicating that the latter
estimate of the ``local'' escape fraction is robust. The extended galaxy
on top of the compact $z\sim 3.3$ source is 3 arcsec in diameter, which
corresponds to 22 kpc if the two galaxies are at the same redshift, while
\cite{bouwens04} found that a typical $L*$ star forming galaxy has an half
light radius of 2 kpc at $z\sim 2.5$.
A plausible explanation for the detection in LyC for source ID=16669 is thus
the presence of a foreground object which mimics the presence of a LyC region
near the $z\sim 3.3$ galaxy.

A similar situation happens for sources ID=52336 and 55594
(Fig.\ref{obj2ima}), where the presence of an extended galaxy and a
compact object, respectively, with relatively bright U band emission
but with significant offset (distance of 2.4 and 1.5 arcsec,
respectively) from the $z\sim 3.3$ objects (green circles) clearly
indicates the superposition of foreground sources. The ``global''
escape fractions for these two galaxies are 140\% and 187\%,
respectively. If we compute instead the ``local'' escape fraction
only from the region with clear detection in the U band, we ended up
with even larger escape fractions, of 966\% and 654\%, respectively.
Even in this case the most plausible explanation is a contamination by
a foreground galaxy prospectively close to the $z\sim 3.3$ source.

\begin{figure*}
\includegraphics[width=18cm,angle=0]{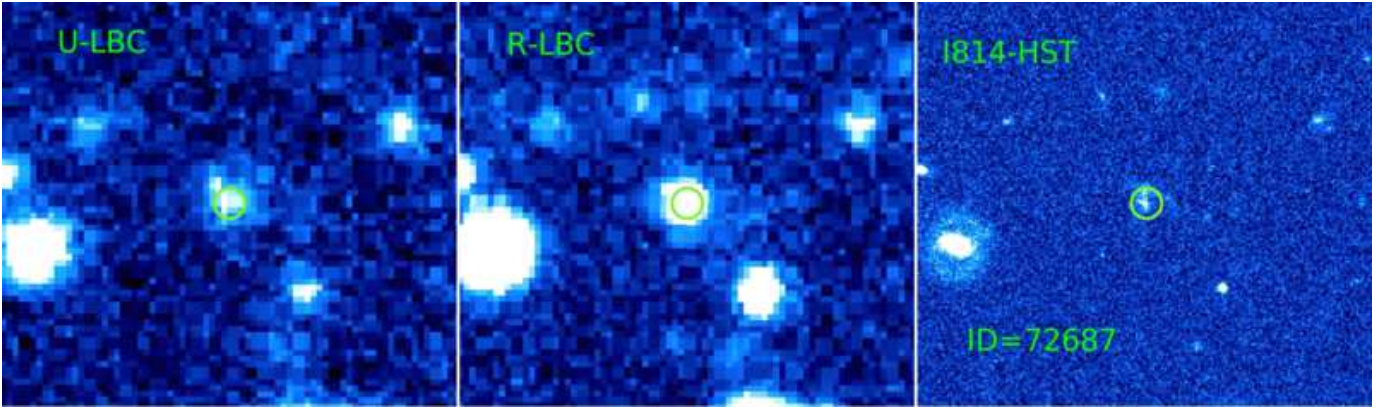}
\caption{The cutouts of source ID=72687 in the U band (left)
and in the R band (central panel) by the LBC instrument.
The right cutout shows the
same sky area observed by HST in the I814W filter. The position of the
spectroscopic redshift is indicated by the green circle.
The size of each image is 14 $arcsec$ by 14 $arcsec$.
In this case the
presence of two distinct objects is not secure, despite the availability
of the HST image, but there is some hints of the presence of two
blended galaxies in the latter. The object in the U band is slightly
offset ($\sim 0.1$ arcsec) from the
center of the source detected in the LBC R and HST I814W bands.
}
\label{obj72687ima}
\end{figure*}

The situation is less clear for source ID=72687 in Table
\ref{table:gal}. This galaxy has an escape fraction of 292\%, clearly
not physical under reasonable assumptions for the stellar populations
involved and for the IGM absorption.
From an analysis of the HST image (Fig.\ref{obj72687ima}) we cannot confirm 
with high confidence the presence of two different objects as in the 
previous cases, although two distinct blobs are seen in the HST 
thumbnail.
From a detailed analysis of the LBC images, we find a small
shift of $\sim 0.5$ pixel, corresponding to $\sim$0.1 arcsec, from the
center of the source in the R band and in the U band. The separation between
the two blobs in the HST image is slightly larger, of 0.3 arcsec.
This could
be an other case of blending of a $z\sim 3.3$ galaxy with a
foreground source, but based on the present data we cannot exclude
that the two blobs detected by HST are all at $z\ge 3$, and this could be an
example of a stellar-powered LyC emission. A detail analysis of the
VUDS spectrum indicates that the redshift of this source is based only
on a noisy Lyman-$\alpha$ and strong CIV and SiIV lines, all in
absorption. There are also other absorption lines not associated with the
$z=3.32$ solution, but it was not possible to derive a secure redshift from
these lines.
Moreover, this source
has strange colors ($u-g=0.13$, $g-K=2.87$ and $K-8.0\mu m=1.94$), which
are not typical of a $z\sim 3$ star forming galaxy. Indeed, the Ultravista
DR1 catalogs by \cite{ilbert13} and \cite{muzzin13} give a photometric redshift
of 0.596 and 0.628, respectively, for this object.
Further analysis
of this very peculiar source will be crucial to understand its nature.

\subsubsection{Galaxies with $f^{rel}_{esc}<100\%$ but detected LyC emission}

In Table \ref{table:gal} there are 6 additional galaxies with detection
in the LyC  wavelengths and estimated values of $f^{rel}_{esc}\le 100\%$.

A careful analysis of the LBC images in the U and the R bands,
complemented by the HST imaging in the I814W band, allows us to
distinguish between real LyC emission and foreground contamination.
Four (ID=19521, 40998, 48734 and 47357) out of these six sources
show a situation similar to the four previously analysed galaxies: the
high spatial resolution images by HST (Fig.\ref{ima4obj})
show the presence of possible faint
contaminants with spatial offsets from the spectroscopically
identified galaxies at $z\sim 3.3$. To check if they are foreground
galaxies, we adopt, as in the previous cases, the test of the ``local''
escape fraction to verify if it is consistent with the emission from
plausible stellar populations and with the IGM attenuation.
The escape fraction has been computed restricting the area to
the region covered by the LyC emission in the U band, instead of computing the
fluxes in the region defined by the
non-ionizing band (in this case the LBC R-band). For all of these
four galaxies, the escape fraction computed ``locally'' on the
U band emission further exceeds the physical limit of 100\%, in some
cases reaching 600\% or more (see notes to Table \ref{table:gal}).

\begin{figure*}
\includegraphics[width=18cm,angle=0]{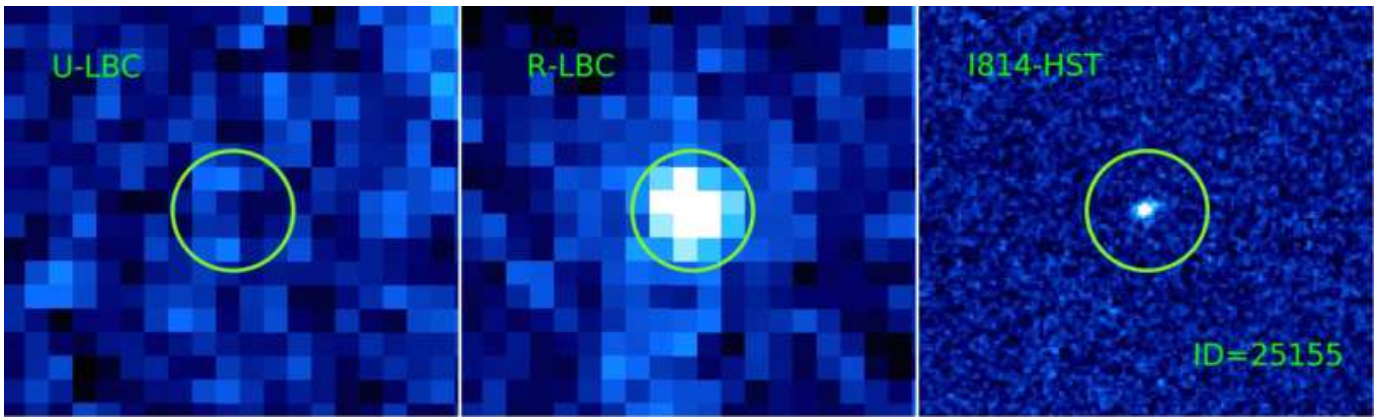}
\caption{The cutouts of source ID=25155 in the U band (left) and in the
R band (center) by the LBC instrument. The cutout on the right shows the
same sky area observed by HST in the I814W filter. The position of the
spectroscopic redshift is indicated by the green circle. The size of the
cutouts is 4.3 arcsec. This object seems
isolated in the HST I814W band, but there is a possible hint for the
presence of a very faint source on the north-west of the bright galaxy
at $z\sim 3.3$. There is a detection at $\sim 2\sigma$ level in the U band
of LBC.
}
\label{obj25155ima}
\end{figure*}

For the remaining two galaxies (ID=25155 and 63327) the presence of a
contaminating foreground is not obvious from HST data, and their ``local''
escape fractions are $<100\%$ if we limit the analysis to the region with
U band emission. The galaxy 25155 (Fig.\ref{obj25155ima})
is isolated and only checking carefully the HST I814W image there is a hint
of a very faint blob slightly offset from the $z=3.3$ galaxy. It has a
distance of 0.1 arcsec from the bright object, thus it is very hard to
draw any conclusions on its true nature. We will assume in the following that
it could be a genuine LyC emitter at $z\sim 3.3$. This galaxy has a relatively
faint magnitude $U=28.6$ and it is marginally detected in the LBC COSMOS
U band image at $S/N=2.2$, with a resulting escape fraction of $\sim 48\%$.
Its optical spectrum by the VUDS survey shows
typical absorption lines of high-z galaxies (Ly-$\alpha$, SiIV, CIV)
so it could resemble the ``Ion1'' source discussed in \cite{vanzella12}
and in \cite{vanzella15}.

\begin{figure*}
\includegraphics[width=18cm,angle=0]{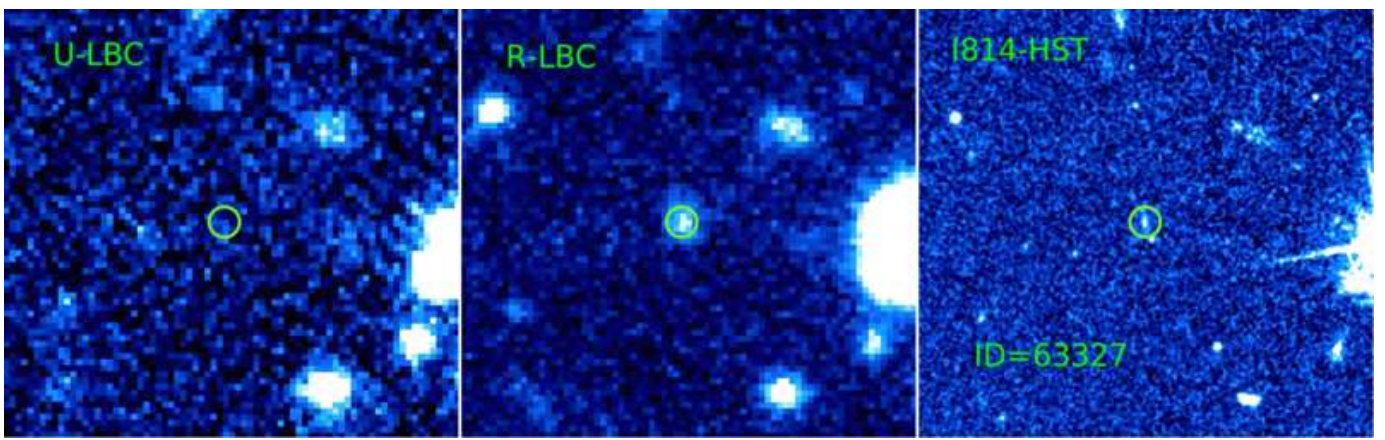}
\caption{The cutouts of source ID=63327 in the U band (left) and in the
R band (center) by the LBC instrument. The cutout on the right shows the
same sky area observed by HST in the I814W filter. The position of the
spectroscopic redshift is indicated by the green circle.
The size of the cutouts is 15 arcsec. In the HST image
there is a faint blob towards the south-west, but it is blended with
the galaxy at $z\sim 3.33$ in the LBC R band. In the LBC U band there is a
very faint emission slightly offset towards the south w.r.t. the R band
position.
}
\label{obj63327ima}
\end{figure*}

In addition, also the detected U-band flux for
galaxy ID=63327 cannot be clearly explained by the presence of a lower
redshift object contaminating the $z\sim 3.3$ source's line of sight.
We have thus another example of a plausible ionizing source
by stellar radiation, with $f^{rel}_{esc}=44.6\%$ and a marginal detection
at $U=28.5$
at $S/N=2.2$. Fig.\ref{obj63327ima} shows this source in the LBC U band
(left), in the LBC R band (center) and in the I814W band by HST (right).
The HST image shows two blobs separated by $\sim$0.7 arcsec,
but even at this high resolution we cannot judge whether this is
a unique object or two distinct galaxies.
There is still the possibility that there is a foreground contaminant
roughly aligned with the $z\sim 3.3$ galaxy making an apparent blend that is
indistinguishable even from HST data.
For further exploring this possibility
we used the high resolution of the HST I814W image to carry out
a de-blended photometry both in the U and in the R band with ConvPhot,
as described in the previous sections, but now adopting the HST image as
a prior. We obtain an upper limit of $U\ge 29.4$ at 1$\sigma$ for the northern
blob, while for the southern region we get a marginal detection $U=28.9$ at
2$\sigma$. Translating the latter value into an escape fraction, we obtain
$f^{rel}_{esc}=119\%$, thus indicating that this can be
a lower redshift interloper,
similar to the ones discussed in the previous section. However, since this
value of the escape fraction can be lower than 100\% for slightly
bluer stellar populations
or a more transparent line of sight, we still consider this object
as a candidate LyC emitter, requiring further analysis in the future.
We have checked also the VUDS spectrum
of this source looking for possible absorption features not consistent
with the
redshift of the $z\sim 3.3$ object, but the S/N is too low to carry out this
test. Its optical spectrum shows only a relatively
narrow Lyman-$\alpha$ in emission, with a symmetric profile.
Deep UV images with HST or deep NIR spectroscopy could finally reveal
its nature, as recently
shown by \cite{siana15} for similar cases in the SSA22 area.

Summarizing, we have found evidence of contamination in 8
out of 45 cases (18\%). This fraction is slightly larger than,
but still consistent with the expected
fraction of contamination (12-15\%) derived following \cite{vanzella10a},
assuming a depth of $U=28.5-29.5$ and a seeing of 1.0 arcsec.
We can thus conclude that there is a high probability that these 8 galaxies
detected in the U band are all foreground interlopers.

%__________________________________________________________________

\section{Results}

\subsection{The escape fraction derived on stacked images}

\begin{figure}
\includegraphics[width=9cm,angle=0]{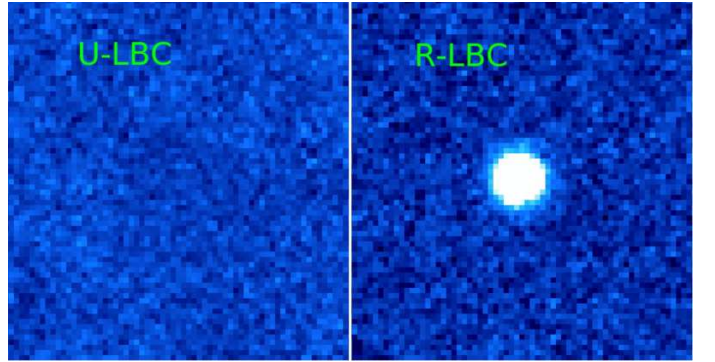}
\caption{The stack of the 37 galaxies with $3.27<z_{spec}<3.40$ in the U
(left) and R bands (right) for the VUDS-LBC/COSMOS, Q0933, and Q1623 fields.
Other objects detected in the R band but not associated with the galaxies
at $z\sim 3.3$ (at the center of the FITS thumbnail) have been masked both in
the U and in the R band image using the segmentation map derived from the
R band. The size of each image is 14 $arcsec$ by 14 $arcsec$.
}
\label{imageUR}
\end{figure}

We derived the stacked images in the U and R bands for the 37 galaxies
presented in Table \ref{table:gal} marked with the label $Cont=No$,
excluding the 8 galaxies where we have
an evidence of possible foreground contamination,
in order to measure the mean escape fraction of the population of star forming
galaxies. We have included in the stack the two galaxies (ID=25155 and
ID=63327) with marginal detection in the U band.
We adopted the same technique described in \cite{boutsia11}. In summary,
we extract a thumbnail of $\sim 45 arcsec\times 45 arcsec$ centered around
the $z\sim 3.3$
galaxies in the U and R band images by LBC, detect all the sources in the
R band and mask all of them in both bands, except for the central target.
Then, we average all the thumbnails of the 37 galaxies using the inverse
square RMS map as a weight. No clipping has been applied during the
stacking. The resulting images (zoomed) are shown in
Fig.\ref{imageUR}.
It is immediately clear the high S/N in the stacked R band image,
given that each individual source is detected at $S/N>10$ in the
original frame. It is also evident that no detection is apparent in
the U band stack.

To be more quantitative, we adopted the same method described in the
previous section for the individual sources, namely we used the
ConvPhot software to carry out a careful evaluation of the depth of
the stacked images. The stacked source in the R band has a mean
magnitude of $R=24.56$ with a $S/N=283$. On the other side, for the U
band we can derive only an upper limit to the emitted flux, reaching
$U=31.4$ at $S/N=1$ (or equivalently $U=30.2$ at $S/N=3$). This
translates into a lower limit for the flux ratio of
$(F_R/F_U)_{obs}=545.1$, corresponding to an upper limit to the
relative escape fraction of $f^{rel}_{esc}\le 2.0\%$ at 1$\sigma$, or
equivalently $6.0\%$ at the 99.7\% confidence level.
If we exclude from the stack the two sources ID=25155 and ID=63327,
i.e. the galaxies detected individually in the U band at $\sim 2 \sigma$ level,
the previous results are not affected at an appreciable level.
The stack of the 35
sources indeed gives a limit to the flux ratio of 549.5 corresponding
to a similar limit for the escape fraction of 2.0\% at 1$\sigma$. Thus the
inclusion of two faint galaxies with marginal detection in the LyC does not
modify the actual limit reached by the stacking.

We split the whole sample in different sub-samples. We explore the
escape fraction for the bright and faint sub-samples. Considering the
11 galaxies with $R\le 24.5$ within the total sample of 37 objects used above,
we end up with a weighted-mean magnitude of $R=23.93$ and an upper limit
for the escape fraction of 2.4\% at 1$\sigma$. The stack of the remaining
26 sources with $R>24.5$ gives instead a weighted mean of $R=24.92$ and
$f^{rel}_{esc}\le 3.1\%$ at 1$\sigma$ (or equivalently $f^{rel}_{esc}\le 9.4\%$ at
3 $\sigma$ level). In this case the slightly worse limit to the
LyC escape fraction is only due to the faintness of the sub-sample.

\begin{figure*}
\centering
\includegraphics[width=14cm,angle=-90]{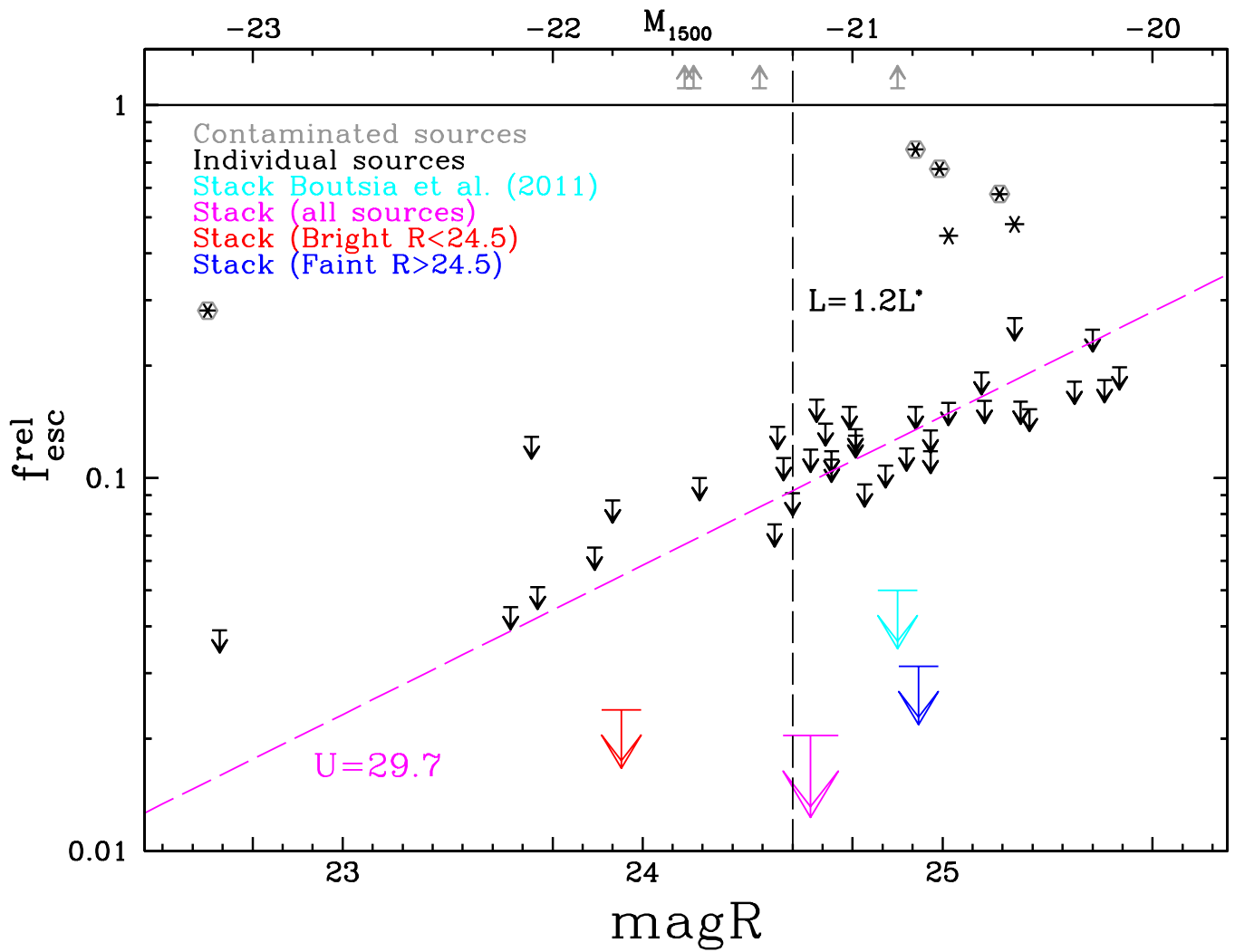}
\caption{
The measured values and the upper limits at 1 $\sigma$ for the LyC
relative escape fraction of galaxies at $z\sim 3.3$ in the COSMOS field. Black
descending arrows show the upper limits associated to individual
galaxies, while the cyan arrow is the limit derived in
\cite{boutsia11} by stacking 11 galaxies. The red arrow is the
limit derived using only relatively bright galaxies with $R\le 24.5$,
corresponding roughly to $L\ge 1.2 L^*$ ($M_{1500}\le -21.2$).
The blue arrow shows the
upper limit for fainter galaxies (stack of 26 objects), while the
magenta one indicates the upper limit for the whole sample (stack of
37 galaxies). The magenta dashed line is the expected $f^{rel}_{esc}$ for a
limit in the U band of 29.7 magnitudes (at S/N=1, assuming an intrinsic
ratio $L_{1500}/L_{900}=3$ and a mean IGM transmission of 0.28). This shows that
the trend of larger escape fractions with fainter R-band magnitudes is
due to the depth of the LBC image in the U band, and it is probably not due to
a physical trend in the population of star-forming galaxies.
Asterisks show the galaxies with detection in the U band and a
significant LyC emission. The grey hexagons and the upper arrows
indicate likely foreground contamination, where the global or the
local escape fractions exceed 100\%. Only two sources (ID=25155 and
ID=63327, black asterisks without the grey hexagons) are an example of
possible LyC emitters.
}
\label{magfesc}
\end{figure*}

Fig.\ref{magfesc} shows the escape fraction at 1 $\sigma$ as a
function of the observed R band magnitudes for all individual sources
and for the stacks.
The trend of larger $f^{rel}_{esc}$ with fainter R band
magnitudes is simply due to the depth of the LBC image in the U
band (U=29.7 AB mag at S/N=1), and it is not due to a physical trend
in the galaxy population (i.e. higher escape fraction for fainter
sources).

The results of the stacks are summarized in Table \ref{table:fesc} and
are also plotted in Fig.\ref{magfesc} with magenta, red, and blue
arrows for the total sample and the bright and faint sub-samples,
respectively. These limits are a factor of 2.5 more stringent than the
previous results by \cite{boutsia11}, and demonstrate the power of
stacking a large sample of galaxies to reach very deep constraints to
the ionizing escape fraction at $z\sim 3.3$.

%__________________________________________________ One column table
\begin{table}
\caption{The escape fraction of stacks}
\label{table:fesc}
\centering
\begin{tabular}{c c c c}
\hline
\hline
Sample & $N_{gal}$ & Rmag & $f^{rel}_{esc}$ \\
       &          & (AB) & (1$\sigma$) \\
\hline
bright ($R\le 24.5$)      & 11 & 23.93 & $\le$0.024 \\
faint  ($R>24.5$)   & 26 & 24.92 & $\le$0.031 \\
\hline
all sources & 37 & 24.56 & $\le$0.020 \\
\hline
\end{tabular}
%add here comments....
\end{table}

\subsection{UV Background from star-forming galaxies at $z\sim 3$}

Using the results obtained in the previous section, we can now derive a
constraint for the UV background produced by the population of star-forming
galaxies at $z\sim 3$. We derived the luminosity density at $900$ {\AA}
rest frame as in \cite{boutsia11}, starting from the analogous quantity
at $1500$ {\AA} rest and correcting it for the intrinsic slope of the spectrum
of these galaxies (adopting $L_{1500}/L_{900}=3$) and the derived limit for
the escape fraction of $2\%$:

\begin{equation}
\rho_{900}^{esc}=\rho_{1500}\cdot f^{rel}_{esc}\cdot (L_{900}/L_{1500})_{int} \, .
\end{equation}

As it can be derived from equation 1,
the product $f^{rel}_{esc} (L_{900}/L_{1500})_{int}$ is equivalent to
$(F_U/F_R)_{obs}exp(\tau_{IGM})$ which are measured quantities
less subject to systematic uncertainties or assumptions.

The only parameter affecting these calculations is the input luminosity
density of non-ionizing radiation, $\rho_{1500}$. To obtain this value it
is sufficient to integrate an observed LF, multiplied
by the absolute luminosity, down to a
specific magnitude limit. We decided to adopt the parametrization of
\cite{rs09} for the LF
at $z\sim 3$ after converting it from 1700 to 1500 {\AA}
rest frame wavelengths, as described in \cite{boutsia11}.
We adopt a slope of $\alpha=-1.73$ and integrate down to an absolute magnitude
of $M_{1500}=-20.2$, which is equivalent to $L\sim 0.5 L^*(z=3)$.
We choose this limit since it corresponds to the limiting magnitude of our
sample, $R\sim 25.5$. The resulting luminosity density is
$\rho_{1500}= 1.27\times 10^{26} erg~s^{-1}~Hz^{-1}~Mpc^{-3}$.
Adopting an upper limit for the escape fraction of $\le 2.0\%$, we obtain
a limit to the ionizing emissivity of $\rho_{900}^{esc}\le 0.85\times 10^{24}
erg~s^{-1}~Hz^{-1}~Mpc^{-3}$ at 1$\sigma$
level, or equivalently $\rho_{900}^{esc}\le 2.55\times 10^{24}
erg~s^{-1}~Hz^{-1}~Mpc^{-3}$ at 3$\sigma$.

We can translate this ionizing luminosity density into a
meta-galactic hydrogen photo-ionization rate
$\Gamma_{-12}$, (ionization rate of HI in unit of $10^{-12} s^{-1} atom^{-1}$)
using the formula by \cite{meiksin09} and \cite{mostardi13}

\begin{equation}
\Gamma_{-12}=\frac{10^{12} \cdot \rho_{900}^{esc} \cdot \sigma_{HI} \cdot
\Delta l \cdot (1+z)^3}{h_P \cdot (3+|\alpha_{UV}|)} \, ,
\end{equation}

where $\sigma_{HI}=6.3\times 10^{-18} cm^2$ is the hydrogen
photo-ionization cross section
at $\lambda=912$ \AA, $\alpha_{UV}=-3.0$ is the spectral
slope of the stellar ionizing radiation
(from \cite{mostardi13}) and $h_P$ is the Planck constant.
The mean free path of ionizing photons, $\Delta l$ is 83.5 proper Mpc at z=3.3,
according to recent estimates by \cite{worseck14a}.
The 1$\sigma$ uncertainty on $\Delta l$ at $z=3.3$ is 4.8 Mpc (i.e. ~6\%).
The uncertainties on the emissivity of galaxies at $1500$ {\AA} rest frame
(due to the uncertainties on the parametrization of the
galaxy LFs) are thus dominating the source of errors on $\Gamma_{-12}$,
as shown in Table 3.
We will discuss in the next section the implications of different
parametrization of the galaxy LFs on the HI photo-ionization rate.

Following \cite{fg09}, at $z\sim 3$ the galaxy radiation can be considered
rather uniform, and in this case the net effect of discrete absorbers is a
hardening of the emissivity, due to the IGM filtering.
As a consequence, a correct value for galaxies can be $\alpha_{UV}=-1.8$ (from
Appendix D of \cite{fg09}), which is in agreement with the value provided
by \cite{hm12} (see their Fig. 10).

\begin{figure}
\includegraphics[width=9cm,angle=0]{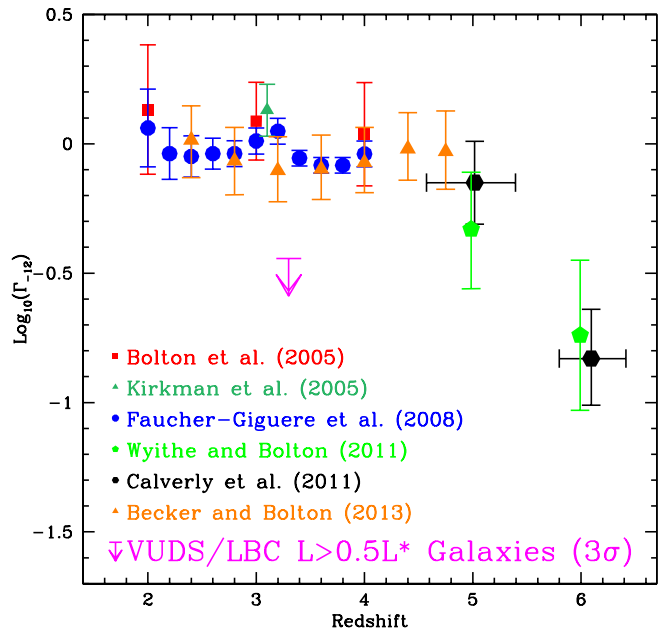}
\caption{The evolution of the UVB, $log_{10}(\Gamma_{-12})$, as a function of
redshift. These measurements are typically derived by an analysis of the
Lyman forest properties or by the proximity effect. The 3$\sigma$ upper
limit from this work ($\Gamma_{-12}\le 0.36$) is indicated by the magenta
arrow. This limit, obtained by star forming galaxies with $L\ge 0.5 L^*$ at
$z\sim 3.3$ shows that the bright galaxies alone cannot be responsible for the
UVB at this redshift.
}
\label{gamma}
\end{figure}

We thus obtain an upper limit $\Gamma_{-12}\le 0.12$ at 1$\sigma$ (or 0.36 at
3$\sigma$) for the ionization rate of relatively bright ($M_{1500}\le -20.2$,
or $L\ge 0.5 L^*$) star forming
galaxies at the mean redshift of our sample (z=3.3).
For comparison, a recent estimate of the ionization rate by \cite{bb13}
gives $\Gamma_{-12}=0.79^{+0.28}_{-0.19}$ at z=3.2,
well above our limit.
Fig.\ref{gamma} shows
the derived upper limit (at 99.7$\%$ c.l. or 3$\sigma$) for $\Gamma_{-12}$
from the VUDS-LBC/COSMOS data,
compared with recent determinations of the UV background (UVB) in the
literature. These estimates are derived through an analysis of the
Lyman-$\alpha$ forest properties (\cite{bolton05,kirkman05,fg08,wyithe11,bb13})
or by the proximity effect (\cite{calverley2011}).
The data shown in Fig.\ref{gamma}, derived from flux decrement analysis in the
IGM Lyman forest of QSO spectra (Bolton et al. 2005; Faucher-Gigu\`ere
et al. 2008; Wyithe \& Bolton 2011), have been scaled to the same
cosmology and to the same IGM temperature-density relation, as
described in Calverley et al. (2011) and in Giallongo et al. (2012).
It is clear from this plot that the ionizing emission of relatively
bright ($L\ge 0.5 L^*$)
star forming galaxies is insufficient to keep the Universe
reionized at $z\sim 3.3$ at more than 99.7$\%$ probability.
It is worth mentioning that the recent determinations of \cite{bb13}
take into account a series of systematic effects on the derivation of
$\Gamma_{-12}$, and thus their error bars can be considered as a robust
estimates for the
uncertainties on the UVB. Our upper limit is clearly lower than their minimum
envelope, giving convincing argument against a major contribution of
bright star forming galaxies to the ionizing emissivity at z=3.
In the next section we will discuss in detail the possible way outs
for the missing HI ionizing photons at $z\sim 3$.

%__________________________________________________________________

\section{Discussion}

Our results are based on two main assumptions, namely the mean
IGM transmission $<exp(-\tau_{IGM})>$ and, in a more subtle way, the adopted
value for the intrinsic ratio $(L_{1500}/L_{900})_{int}$. We will discuss in the
following sections the implications of the choice of these two
parameters on our conclusions.

\subsection{The variance of the IGM and ISM transmissions}

In the previous sections we have adopted a mean value of 0.28 for the
mean IGM transmission close to the Lyman limit. This value however is
subject to large cosmic variance due to differences on the absorption patterns
in various lines of sight, especially due to the patchy nature of the
Lyman limit systems (LLS) and Damped Lyman-$\alpha$ systems (DLAs),
which are able to block the path of
ionizing photons through the IGM.
The IGM absorption at $\sim 900$ {\AA} is thus very
stochastic because it is related to the probability of intercepting
dense absorbers, with $log(N_{HI})\ge 17$.

We have estimated the variance of the
IGM transmission $exp(-\tau_{IGM})$ at $\sim 900$ {\AA} rest frame adopting the
Monte Carlo simulations of \cite{inoue08} and the recent updated
absorber statistics by
\cite{inoue14} used also by \cite{vanzella15}.
Recently, \cite{thomas15} estimated the same quantity using deep VUDS
spectra of galaxies in the redshift interval $2.5\le z\le 5.5$,
finding an excellent agreement with the IGM transmission obtained from
quasars' lines of sight at $z\le 4$.
We considered 10,000 different realizations of the IGM transmission
at $z\sim 3.3$ and convolved each of them with the profile of the LBC U band
filter, thus obtaining a mean transmission and its variance. We obtained
a mean value of
$<exp(-\tau_{IGM})>=0.25$, similar to the adopted value of 0.28 based on the
\cite{worseck14a} IGM transmission, and a dispersion of $\sim 0.2$.
The resulting distribution of IGM transmission at $z\sim 3.3$ is
asymmetric, with a tail toward low $exp(-\tau_{IGM})$
due to the presence of DLAs and LL systems. A fraction of $11\%$ of
all the simulated lines of sight
shows a transmission $exp(-\tau_{IGM})$ less than 0.01, thus it is important
to have a large number of galaxies with measured LyC escape fraction
in order to minimize
the stochastic effect of intervening absorbers.
Considering our case of 37 different lines of sight, the dispersion of
$<exp(-\tau_{IGM})>$ is 0.03.

Another source for the skewed probability distribution function towards low
values of the escape fraction of galaxies is the patchiness of the
ISM, which has been recently explored by
\cite{cenkimm15}. Using high-resolution large scale simulations with radiative
transfer calculations, they show that LyC photons can escape from a galaxy
through small solid angles with complete transparency, while the majority of
the remaining LoS are opaque. Adopting their simulations at $z=4$ and at
a virial mass $10^{10}\le M\le 10^{11}M_{\odot}$, we estimate that for an input
escape fraction of 2\% with 37 galaxies the confidence interval at
1 $\sigma$ level is between $f^{rel}_{esc}=1.4\%$ and 2.7\%.
Simulations are not available
for our redshift and virial mass range, but the trend with redshift and masses
derived by \cite{cenkimm15} shows that our extrapolation at $z\sim 3.3$ and
higher luminosities should be fair enough.
Combining together the uncertainties due both to the stochasticity of the
IGM and the variance of the ISM in different LoS,
the confidence interval at 1 $\sigma$ is between $f^{rel}_{esc}=1.3\%$ and 2.8\%.

Thus we can conclude that with our relatively large sample of galaxies we
are able to limit the variance due to both IGM transmission and patchy
ISM of the star forming galaxies. We can exclude that the lack of ionizing
photons we observed in the VUDS-LBC/COSMOS sample is due to the low number
statistics and variance in the IGM or ISM transmission.
In the near future, the combination of HST imaging, deep UV observations and
large spectroscopic surveys will provide more than 100 galaxies where these
measurements can be repeated.

\subsection{The intrinsic ratio $(L_{1500}/L_{900})_{int}$}

Our choice of the $(L_{1500}/L_{900})_{int}$ ratio does not enter
directly in the evaluation of the total ionizing emissivity since the
dependencies of $\rho_{900}^{esc}$ from the intrinsic ratio
$(L_{1500}/L_{900})_{int}$ is canceled out, once equations 1 and 3 are
combined. As we have already discussed, in fact, the value of the HI
photo-ionization rate does depend only on the observed ratio between
the U and the R band photometry and on the IGM correction. However,
the $(L_{1500}/L_{900})_{int}$ ratio can be used to provide us a with
a physical interpretation of the amount of UV ionizing photons
escaping from a galaxy, and enters in our definition of the sample. We
will discuss this latter aspect here.

The relative escape fraction derived for our sample of star forming galaxies
assumes a given ratio between the intrinsic luminosity at 1500 and 900 {\AA}
rest frame, here taken to be $(L_{1500}/L_{900})_{int}=3$.
This value has been adopted for a direct comparison with previous works
dealing with similar subjects (\cite{steidel01,shapley06}).
The intrinsic ratio $(L_{1500}/L_{900})_{int}$
depends on the star formation history, stellar populations, age, metallicity,
initial mass function of the investigated galaxies.

\cite{vanzella10b} showed that the intrinsic value
$(L_{1500}/L_{900})_{int}$ can be as large
as 7 if considering mean values for the physical parameters of high-z
star-forming galaxies. Adopting this new value would approximately
double the escape
fraction we measured for the VUDS-LBC/COSMOS sample, but would not
change our conclusions on their contribution to the ionizing emissivity.
Moreover, adopting a
value of 7 for $(L_{1500}/L_{900})_{int}$ would result in even higher
values for the relative escape
fraction of the
galaxies which already show $f^{rel}_{esc}\ge 100\%$, thus strengthening
our conclusions about their nature of low redshift interlopers.

Alternatively, \cite{topping15} adopted a recent rendition of stellar
evolutionary tracks as input in their galaxy synthesis model, finding a 50\%
increase in the LyC production efficiencies w.r.t. previous calculations.
Changing the intrinsic value of $(L_{1500}/L_{900})_{int}$ to 2 would not change
our conclusions on the ionizing emissivity, as discussed above.
Moreover, since the relative escape fraction we derived with the ``local''
method is significantly greater than 100\% (175-966\%) for all
the galaxies that were considered to be affected by foreground
contamination, a correction of
50\% downward would only partially alleviate the tension for one object
(ID=19521), but would not solve the problem for the other galaxies.

We can thus conclude that the assumed value of $(L_{1500}/L_{900})_{int}=3$
is not significantly affecting our conclusions on the ionizing emissivity of
star forming galaxies at $z\sim 3.3$.

\subsection{Comparison with other results}

According to \cite{cooke14}, the LBG selection could be biased against
LyC sources, since a galaxy with significant emission at $\lambda\le
912$ {\AA} rest-frame would avoid the classical drop-out selection. We
show here that our sample is not affected by these systematics.
In fact, even assuming an escape fraction of
100\% and a completely transparent IGM (very unlikely), the typical
color of a galaxy at $z\sim 3$ would be $U-R\ge 1.2$, above the color criteria
normally adopted to select candidate galaxies for spectroscopic follow
up (see for example \cite{vuds} and \cite{boutsia14}). A recent paper
(\cite{vanzella15}) confirms our conclusions with state-of-the-art
simulations adopting the updated recipes of \cite{inoue14} for the
IGM transmission.

The result found from the VUDS-LBC/COSMOS sample can be compared with recent
achievements on the detection of LyC emitters at $z\ge 3$.
\cite{iwata09,nestor11,nestor12} and \cite{mostardi13} found a
significant number of ionizing sources in their surveys. In the SSA22
area, \cite{iwata09} found evidence for $f^{rel}_{esc}>4\%$ for the whole
population of 73 LBGs analysed, but provided also evidence for a significant
spatial offset between the ionizing emission and the
non-ionizing one for some of them.
Moreover, there are few objects detected in LyC with a flux ratio
$F_{1500}/F_{900}$ smaller than the typical values expected from PopII
stellar population with standard Salpeter IMF (or other non top-heavy
ones), even assuming a transparent IGM. Moreover,
they didn't carry out a local
estimate of the escape fraction in the position of the detected LyC
sources.

On the same SSA22 field, \cite{nestor11} and \cite{nestor12} show nine
LBGs and 20 Lyman-$\alpha$ emitters (LAEs) with LyC detection out of a
sample of 41 LBGs and 91 LAEs (all spectroscopically confirmed). They
started from a different narrow band image centered at $\sim 3640$
{\AA} which is deeper than the one used by \cite{iwata09}, at $\sim
3590$ \AA. A careful analysis of their LBG detections, however, shows
that even in this case the LyC emission for many $z\sim 3$ sources is
offset by 0.4-1.0 arcsec. The observed ratio between the $900$ and
$1500$ {\AA} rest frame emission is difficult to reconcile with what
expected by standard stellar populations (\cite{vanzella12}). The HST
images in I814W filter available for few of them show the presence of
clearly separated galaxies, sometimes fainter at $1500$ {\AA} than in
the ionizing continuum (i.e. their C16 object), with a resulting
escape fraction well exceeding 1000\% if estimated in the LyC
position. For the majority of them no HST imaging is available but
even ground-based images often show the presence of slightly offset
emission in LyC w.r.t. the non ionizing continuum. Similar conclusions
can be reached for the \cite{mostardi13} sample where they adopt the
same analysis as in \cite{nestor12}. They found at $z\sim 2.8$ four
LyC emitters out of 49 LBG galaxies and 7 LyC emitters out of 91
Lyman-$\alpha$ emitters. In this case the lack of high spatial
resolution data from HST for the majority of the sample prevents any
detailed analysis about possible contamination by
interlopers/foregrounds. These conclusions have been strengthened by
recent observations by \cite{siana15}, where they found no convincing
detection in their deep HST imaging with UVIS of 5 LyC emitters
extracted from the sample of \cite{nestor11}, or by \cite{mostardi15},
where only one robust LyC emitter has been found after a re-analysis of
a sample of 16 galaxies by \cite{mostardi13}. More interestingly, the only
robust candidate LyC emitter by \cite{mostardi15}, the galaxy MD5b,
has an observed ratio $F_{UV}/F_{LyC}=4.0\pm 2.0$, equivalent to a relative
escape fraction of 75\%, assuming complete transmission of the IGM.
If instead a mean value of $<exp(-\tau_{IGM})>=0.4$ at $z\sim 3.1$
will be adopted, following \cite{inoue14}, the relative escape fraction
of MD5b turns out to be 188\%.
Imposing the constraint of a physical value for the relative escape fraction
of $f^{rel}_{esc}<100\%$, the line of sight of MD5b must be very transparent
$exp(-\tau_{IGM})>0.75$, which corresponds to a probability $<10^{-4}$,
according to \cite{inoue14}.
This could be an indication that also this
galaxy is a low-z contaminant, similar to the other cases studied
by \cite{mostardi15}.

If we compare the apparent statistic of LyC detections in VUDS, we
have 10 individual detections out of 45 galaxies (22\%), similar to the
value by \cite{nestor11} but significantly more than the value by
\cite{mostardi13} (8\%), probably due to their shallow narrow band
imaging.

In summary, it is difficult to compare in a homogeneous way different
results obtained through imaging of different quality in terms of depth and
angular resolution. It appears however that the apparent discrepancies
from different databases tend to converge toward low values for the
average escape fraction if we remove from the analysis the probable low-z
contaminants, i.e. the objects
with flux ratios $F_{900}/F_{1500}$ significantly higher than that
predicted by spectral synthesis models,
where a spatial offset between $F_{900}$ and $F_{1500}$ is present
(\cite{vanzella12,siana15}).

\subsection{The sources of ionizing photons at $z\ge 3$}

In the previous section we have shown that bright ($L\ge 0.5 L^*$)
galaxies at $z\sim 3.3$ cannot provide alone the required LyC emissivity
($\Gamma_{-12}<0.36$ at 99.7\% confidence level) to explain the
measured UVB. The AGN contribution at $z=3.3$ is $\Gamma_{-12}=0.37$
(from Fig. 8a of Haardt \& Madau 2012). The measured value at $z=3.2$
is $\Gamma_{-12}=0.79_{-0.19}^{+0.27}$ from Becker \& Bolton (2013).
Even considering the uncertainties, there is a tension between the
measured value of $\Gamma_{-12}$ and the summed contributions of AGNs
plus bright galaxies at $z=3.3$. This implies that an additional
source of ionizing photons ($\Gamma_{-12}=0.06-0.33$) is probably required in
order to reach the observed value of the photo-ionizing background,
taking into account its uncertainties.
Looking for ionizing sources among the fainter galaxy population is
the obvious next step. The faint population is more numerous and
could show a higher average escape fraction
(\cite{rsl10,aft12,fontanot13,pkdv13,kimmcen14,wise14,paard15,roy15,dc15}).
In the following, we have explored different combinations of the
escape fraction, slope $\alpha$ of the LF and
magnitude limit $M_{1500}$, adopted to compute the ionizing
emissivity, looking for the sources responsible for the observed UVB
at $z\sim 3.3$. A summary is provided in Table \ref{table:lfgamma}.

Adopting quite steep LFs as derived by \cite{rs09}
with a power-law index $\alpha=-1.73$, and integrating down to
$M_{1500}=-17.2$ (or $R=28.5$), we obtain an ionization rate of
$\Gamma_{-12}\le 0.38$ at 1 $\sigma$, assuming a constant escape
fraction of 2\%, as for the brighter galaxies.

The relative escape fraction of galaxies brighter than $M_{1500}\sim -20.2$
($L=0.5 L^*$) should be $\sim 16$\% in order to provide a
photo-ionization rate $\Gamma_{-12}\sim 0.95$, in agreement with the
observed value at $z\sim 3$ shown in Fig.\ref{gamma}. Alternatively,
the escape fraction can remain of the order of few percent but the LF
must be integrated down to very faint magnitudes, at $M_{1500}=-14.2$
(\cite{fontanot13,alavi}) to have $\Gamma_{-12}\sim 0.9$.
The calculations above are quite
conservative, since we are assuming a rather steep LF ($\alpha=-1.73$).
At $z\sim 3$ the estimated slopes of the
galaxy LF
are generally flatter, with $\alpha\sim [-1.4,-1.5]$
(\cite{st06,finkelstein14,parsa15}), with the exception of the LF of
Lyman-$\alpha$ emitters by \cite{cassata11} that has $\alpha\sim -1.8$
at $z\ge 3$. Adopting the parametrization of \cite{st06}, with
$\alpha=-1.43$, and integrating down to $M_{1500}=-14.2$, the ionizing
emissivity is $\Gamma_{-12}\le 0.31$, insufficient to keep the
Universe reionized. Similarly, if we adopt the LF by
\cite{cucciati12} with a slightly steeper slope ($\alpha=-1.50$), but
brighter $M^*$, we find an emissivity of $\Gamma_{-12}\le 0.26$ at
$M_{1500}=-14.2$, far from providing enough ionizing photons to maintain
the Universe reionized. The latter LF is based on the VVDS (\cite{vvds})
spectroscopic survey, so it should be robust against interlopers in
the $z\sim 3$ LF. At $z\ge 3$, \cite{cassata11} find that
the LF of LAEs is relatively steep, with a
slope of $\sim -1.8$. However, these galaxies are much fainter than the typical
UV selected galaxies discovered at the same redshifts. They measured a star
formation rate density which is a factor of 5 lower than the one by
\cite{rs09}, with the implication of a lesser contribution of the LAEs to
the HI ionizing background at $z\sim 3$. Table \ref{table:lfgamma} summarizes
the different HI photo-ionization rates derived for different parametrization
of the $z\sim 3$ galaxy LF and at different luminosities.

%__________________________________________________Two column table
\begin{table*}
\caption{The HI photo ionization rate for different LF parametrization
and integration limits}
\label{table:lfgamma}
\centering
\begin{tabular}{c c c c c c}
\hline
\hline
$f_{esc}^{rel}$ & LF & $\alpha$ & $\Gamma_{-12}(M_{1500}\le -20.2)$ &
$\Gamma_{-12}(M_{1500}\le -17.2)$ & $\Gamma_{-12}(M_{1500}\le -14.2)$ \\
   &  &  & $L\ge 0.5L^*$ & $L\ge 0.03L^*$ & $L\ge 0.0016L^*$ \\
\hline
$<$2\% & \cite{rs09} & -1.73 & $<0.12$ & $<0.38$ & $<0.93$ \\
$<$2\% & \cite{st06} & -1.43 & $<0.10$ & $<0.23$ & $<0.31$ \\
$<$2\% & \cite{cucciati12} & -1.50 & $<0.10$ & $<0.19$ & $<0.26$ \\
$<$2\% & \cite{cassata11} & -1.78 & - & $<0.08^{a}$ & - \\
21\% & \cite{rs09} & -1.73 & 1.24 & 3.94 & 9.73 \\
\hline
\hline
\end{tabular}
\\
The upper limits on $f_{esc}^{rel}$ and $\Gamma_{-12}$ are at 1$\sigma$.
(a) \cite{cassata11} found a star formation rate density of
$0.008 M_{\odot}/yr/Mpc^3$ integrating the LAE LF down to $0.04L^*$.
This SFR density is a factor of 5 below the one by \cite{rs09}.
\end{table*}

As a comparison with other works
(e.g. \cite{iwata09,nestor11,mostardi13}), we have carried out the
stacking of all the 45 sources in Table \ref{table:gal}, regardless of
the contamination by possible foreground sources. In this case we have
a clear detection in the U band with $S/N=11$, of $U=28.87$. This
translates into an escape fraction of $f^{rel}_{esc}=21\%$, in agreement with
previous findings of \cite{iwata09,nestor11,mostardi13}. The resulting ionizing
emissivity of $L\ge 0.5 L^*$ galaxies (or equivalently $R\le 25.5$, our
limit) is $\Gamma_{-12}=1.24$, thus all the LyC photons measured in the
Lyman forest should be provided by relatively bright galaxies, with the
consequence of a negligible contribution by fainter galaxies or by AGNs.
Taken at face value, this would imply that only
galaxies brighter than $\sim 0.5 L^*$ are contributing to the ionizing
radiation at $z\sim 3.3$, thus excluding fainter galaxies and the QSO/AGN
population as sources of ionizing photons. This is at variance with the
conclusions of \cite{lusso15} and \cite{glikman11}, who found that bright
QSOs at $z\sim 2.4$ and at $z\ge 4$, respectively, should provide at least
$\sim 50\%$ of the HI ionizing photons.

If we integrate
the \cite{rs09} LF down to $M_{1500}=-17.2$,
corresponding to a limit of $R=28.5$, and compute the escaping
emissivity at $900$ {\AA} rest-frame assuming $f^{rel}_{esc}=21\%$,
we end up with $\Gamma_{-12}=3.94$,
well above the measured value at $z=3.3$ shown in
Fig.\ref{gamma}.
We can thus conclude that a large fraction of the galaxies with
detected LyC emission in the VUDS-LBC/COSMOS field are plausibly
contaminated by foreground interlopers, as discussed above and also
shown by \cite{siana15} and \cite{mostardi15}.
In this case, star-forming galaxies alone will probably not be
able to keep the Universe ionized at $z\sim 3.3$. The situation can be
different at higher redshifts.

Recent evaluations by \cite{finkelstein14} show a steepening of
the faint end of the LF with a slope as steep as -2
at $z\sim 6-7$, i.e. close to the epoch of reionization. Very high
redshift star forming galaxies down to $M_{1500}=-17$ at $z\sim 7$
could ionize the IGM if they had an escape fraction of $\sim
13$\%. Similarly, \cite{bouwens15} modeled the redshift evolution of the
UV background integrating the LF of star-forming galaxies
down to $M_{1500}\sim -13$ and assuming an escape fraction of 0.10-0.40.
These parametrization allow them to reproduce the observed trend
of the ionization rate $\dot{N}_{ion}$ inferred at $z>6$ by a number of
independent observations (see their Table 1).
Using the same approach, \cite{dc15} concluded that star-forming galaxies
at $z>6$
down to $M_{1500}\sim -15$ and with $f_{esc}\sim 0.16-0.42$ could complete
and maintain the reionization process.
Given our limits ($f^{rel}_{esc}\le 2\%$) at $z\sim 3.3$, an increase
in the escape fraction should be invoked for star-forming galaxies with
lower luminosities
and/or increasing redshifts to make these sources the major
contributors to the reionization at $z>6$.
The observational evidence that galaxies at higher redshifts and fainter
luminosities are bluer and less dusty than bright and low-z galaxies
(\cite{dc15}) and the recent
finding that ionization parameter can be higher for $z>6$ galaxies
(\cite{stark15}), may indicate that high-z galaxies can be
more efficient ionizing agents than galaxies at $z<3$.

The alternative possibility that the AGN population can provide most
of the ionizing flux in the Universe cannot be excluded,
however. Faint multiwavelength surveys based on very deep near
infrared data and at the same time ultra-deep X-ray images are
discovering a population of faint AGNs at very high redshifts ($z\sim
4-7$) that could provide the required ionizing photons (e.g. Fiore et
al. 2012, Giallongo et al. 2012, 2015).
The main uncertainty even for
this population relies on the escape fraction of ionizing photons from
their host galaxies.
It is not clear whether the AGN escape
fraction remains high for progressively fainter sources,
but, at variance with the galaxy population,
the brighter AGNs and QSOs having $f^{rel}_{esc}\sim 100\%$ are known
to ionize their neighborhood even at $z\sim 6$ (\cite{worseck14a}).
In the local Universe, recent results by \cite{stevans14}
seem to indicate that even the faint AGN population has an escape
fraction close to unity. Accurate measurements of the escape fraction
of faint AGNs at $z>3$ will be important to confirm this hypothesis.

\subsection{Comparison with theoretical predictions}

Theoretical predictions of the ionizing photon leakage from star-forming
galaxies at high-z are far from being settled, despite the large
effort put in complex numerical simulations, and different conclusions
have been obtained by various authors on this subject in the past.

Using high resolution simulations, \cite{gnedin08} derived an absolute escape
fraction of $\sim 1-3\%$ for galaxies with mass $M\ge 10^{11} M_\odot$
and star formation rate $\sim 1-5 M_\odot/yr$ at $3<z<9$. This model
is in slight agreement with our result on the VUDS-LBC/COSMOS region. They
predict also that the escape fraction declines steeply at lower masses
and SFR, while they found large variance among different lines of
sight within individual galaxies. According to their model, a high-z
galaxy can be more transparent along the galactic poles than through the
galactic disk. They explain the low values of escape fractions with a
small fraction of young stars located just outside the edge of the HI
disk. In lower mass galaxies, this disk is thicker and more extended
relative to the distribution of young stars compared to massive
galaxies, causing a progressively lower escape fraction.

Differently from \cite{gnedin08}, the simulations of \cite{yajima14}
found that $f^{abs}_{esc}$ remains roughly constant at a value of 20\%
(corresponding to $f^{rel}_{esc}=52\%$ for $E(B-V)=0.1$ and a Calzetti
extinction law) from z=0 to z=10, and it does not show any clear
dependence on the galaxy properties (i.e. luminosity). This is at
variance with our results discussed in previous sections
($f^{rel}_{esc}\le 2\%$ at $L\ge 0.5 L^*$).

Using ray tracing simulations with spatial resolution of 3.8 pc,
\cite{jihoon13} find $f^{abs}_{esc}=1.1\%$ for a spiral galaxy in a
dark matter halo of $\sim 2\times 10^{11} M_{\odot}$. They do not predict
the dependence of the escape fraction from galaxy mass or luminosity.
Models based on hydrodynamical simulations or
radiative transfer calculations (\cite{rsl10,pkdv13,kimmcen14,wise14})
generally predict an increase of the escape fraction at fainter luminosities
or at lower halo masses, with a large value of the absolute escape fraction
($\sim 15-80\%$) at $M_{1500}\sim [-14,-10]$ and at high redshifts ($z>4$).

Only recently, however, the predictions from cosmological
hydrodynamic simulations start converging on a general consensus 
that the escape fractions is increasing towards low masses/luminosities.
For example \cite{paard15}, using high-resolution cosmological
hydrodynamic simulations of galaxy formation, find that only in 30\%
of the halos with $M_{200}=10^8 M_\odot$ (corresponding to a stellar mass
of $\sim 10^5 M_{\odot}$; see their Fig. 5, central panels) the absolute escape
fraction is higher than 1\% at $z>9$, while $f^{abs}_{esc}\ge 10\%$ at
z=6 and/or at lower masses, with a strong anti-correlation between
escape fraction and halo masses. Again, the population of ultra-faint
galaxies in their model is essential in providing the LyC photons
needed to sustain reionization.
Our VUDS-LBC/COSMOS survey is limited to much brighter objects
($L\ge 0.5 L^*$), thus we cannot
provide meaningful constraints to those models.
Using 2-D hydrodynamic simulations, \cite{roy15} find
$f^{abs}_{esc}\sim 5\%$ for Milky Way (MW) galaxies, with a trend of increasing
the escape fraction for lower mass galaxies ($f^{abs}_{esc}\sim 10\%$
for galaxies ten times less massive than the MW).

Using high resolution numerical simulations (0.1-4 pc) at $z\ge 6$,
\cite{ma15} find a large time variability of $f_{esc}^{abs}$
(0.01-20\%), with a mean value $\le 5\%$ over long timescales, without
any strong dependence on mass or redshift. They also show that models
based on ``sub-grid'' simulations, which are not able to resolve the
star formation processes in dense molecular clouds, usually
over-predict the HI ionizing escape fraction.

Phenomenological models (\cite{hm12,kuhlen12,aft12,fontanot13})
explored different parametrization of the escape fraction
of high-z galaxies to explain the evolution of the UV
background and the Thomson optical depth measured at that time by
WMAP. They all assume a low escape fraction at $z\sim 3$, of the order
of $f_{esc}^{abs}\sim 3-5\%$, but they require a strong evolution with
luminosity or with
redshift, in order to fit the higher redshift data. The escape
fraction of VUDS galaxies at $z\sim 3.3$ is still compatible ($\le 6\%$ at
3$\sigma$ level) with these values. More stringent limits and the
exploration of the escape fraction at lower luminosities and/or higher
redshifts can thus provide significant constraints to these models.

%__________________________________________________________________

\section{Summary and Conclusions}

We have analysed the escape fraction of 45 galaxies selected in the
COSMOS region (\cite{scoville}) with robust spectroscopic redshifts
between $3.27\le z\le 3.40$ from the ultra-deep VIMOS spectroscopic
catalog provided by the VUDS survey (\cite{vuds}). Deep U and R band
images obtained by LBC at the prime focus of the LBT telescope
(observations described in \cite{boutsia14}), reaching 1$\sigma$
magnitude limits of $U\simeq 29.7$ and $R\simeq 28.2$, were used to
measure the ionizing escape fraction from the galaxy sample down to a
few percent level. The main results are the following:

\begin{itemize}
\item
A sub-sample of ten galaxies with $R\le 25.5$ has a measured U band
detection that corresponds to a LyC emission of $f^{rel}_{esc}\ge
28\%$. A detailed analysis of their properties has been performed
computing the ''local'' escape fraction in the spatial region of the U
band where the supposed LyC flux was measured. Unphysical escape
fractions $f^{rel}_{esc}\sim 300-1000\%$ have been derived for eight
out of ten objects. The relative $f_{esc}$ of these galaxies remains
above 100\% even assuming a transparent IGM and non standard stellar
populations/IMFs (PopIII or top heavy IMF).
\item
High resolution HST images in the I814W band suggest that the supposed
LyC emission comes usually from distinct interlopers slightly offset
from the positions of the star-forming galaxies at $z\sim 3.3$. It
turns out that eight out of the ten $z\simeq 3.3$ galaxies show
detection in the LBC U band contaminated by light from foreground
galaxies at $z<3$. Only two galaxies could be true LyC emitters,
though the significance of their detection in the U band is marginal,
at 2$\sigma$ level only, and for one of these two the possibility of
contamination from an interloper is still open. Further observations
of these peculiar sources will be fundamental to define their physical
nature.
\item
We stacked the U and R band images of the remaining clean sub-sample
of 37 galaxies at $z\sim 3.3$ (including the two marginal LyC detections)
to compute their average ionizing escape
fraction. The stacks are undetected in the U band down to $\sim
31.4$ magnitudes (at 1 $\sigma$). This corresponds to an upper limit
of $f^{rel}_{esc}\le 2\%$ (1 $\sigma$) at $R\le 25.5$ or equivalently
$L\ge 0.5 L^*(z=3)$.
\item
Considering the stochasticity of the IGM and the patchiness of the ISM
along different LoS as
two sources of uncertainties for the derivation of the escape fraction, the
relatively large number of galaxies analysed here, i.e. 37, allows us to
constrain $f^{rel}_{esc}$ between 1.3\% and 2.8\% at 1 $\sigma$ confidence
level.
\item
Integrating the \cite{rs09} $1500$ {\AA} LF at $z\sim
3$ down to $0.5 L^*$ we have derived a significant upper limit to the
ionizing emissivity and photo-ionization rate of relatively bright
star-forming galaxies $\Gamma_{-12}\le 0.12$ at 68\% confidence level
(corresponding to $\Gamma_{-12}\le 0.36$ at 99.7\% confidence
level). This limit is a factor of two lower than the recent estimates
derived from the IGM observed in the Lyman-$\alpha$ forest spectrum of
bright QSOs at z=3, as shown in Fig.\ref{gamma}. We can thus conclude
that galaxies brighter than $0.5 L^*$ give a small contribution to the
UV background needed to keep the Universe ionized at $z\sim 3$.
\item
Our results differ from those obtained at similar redshifts by
\cite{iwata09,nestor12,mostardi13}. The reason resides mostly on the
interpretation and quality of the database used. Robust spectroscopic
redshifts are available for all the galaxies of our sample and high
resolution HST images supported by spectral synthesis models have been
used to clean our sample from bona-fide low redshift interlopers.
Only few objects are currently genuine LyC emitters, and further analysis
is important to understand their nature. 
Similar conclusions
have been obtained by \cite{vanzella12} and \cite{vanzella15}
in the GOODS-South field, by
\cite{siana15} in the SSA22 area, and by \cite{mostardi15} in the HS1549+1933
field.
Future NIR spectroscopic follow up of all the faint interlopers can assess
their nature.
\end{itemize}

In summary, considering that the expected AGN contribution at $z\sim
3$ amounts to $\Gamma_{-12}=0.37$ (Haardt \& Madau 2012), an
additional contribution of $\Gamma_{-12}\sim 0.06-0.33$ is required to
match the value of $\Gamma_{-12}=0.79^{+0.28}_{-0.19}$ observed by
\cite{bb13}. This additional contribution could be provided by a
faint ($L<0.5L^*$) galaxy population with an escape fraction of
$f^{rel}_{esc}\ge 6-20\%$ (the exact value depends on the assumed
slope of the galaxy LF), which is at least three times larger than our
upper limit derived from the brighter population.

At higher redshifts ($z=4-7$), where the contribution of bright AGNs
to the photo-ionization rate is expected to be progressively less important
(\cite{hm12}), the escape fraction of
star-forming galaxies is expected to be larger than the one measured
here at $z\sim 3$. The galaxy LF at $z>6$ is
quite steep ($\alpha\le -1.7$), compared to the LFs at $z\sim 3$
(e.g. \cite{cucciati12,finkelstein14}), and the contribution of faint
galaxies (down to $M_{UV}\sim -14$ with $f^{rel}_{esc}\sim 10-20\%$) is
required to provide enough LyC photons to effectively contribute to
the reionization of the Universe. This hypothesis however would imply
a rapid change of the escape fraction of ionizing radiation from
star-forming galaxies significantly fainter than $L^*$, which is
difficult to explain in the current theoretical scenarios for galaxy
formation and evolution (\cite{ma15}). From the observational point of
view, instead, there are indications that galaxies are bluer at high
redshift and faint luminosities, thus plausibly implying younger and less
absorbed stellar populations, with possible larger contribution of
ionizing photons than the ones present at $z\sim 3$
(\cite{stark15,bouwens15,dc15}).
A significant contributions to the reionization by different
populations is however still possible, since the escape fraction of
galaxies at $z=4-7$ is currently a free parameter. Faint AGNs are the
most plausible competitors (\cite{giallongo12,worseck14b,giallongo15})
but more exotic alternatives (primordial cosmic rays, PopIII stars
forming high-mass x-ray binaries or Hawking evaporation of primordial
mini black holes) could also play a role (\cite{tueros14,belotsky14}).

In the future, different activities can be pursued to shed light on
the topic of the escape fraction of high-z galaxies and their relation
with the reionization of the Universe. In particular, we are looking
for possible faint ionizing galaxy populations combining high spatial
resolution HST images from the CANDELS survey in GOODS-North with deep
LBC imaging in the U band (\cite{grazLBC}) in order to measure the ionizing
escape fraction at $z\sim 3$ in galaxies (and AGNs) fainter than those
included in
the present VUDS-LBC COSMOS field (i.e. $R\ge 25.5$). At relatively
bright luminosities, the present sample can be significantly enlarged
by the other two fields of the VUDS survey, i.e. the VVDS 2-hour
region and the ECDFS, which benefit respectively from relatively deep
UV images by CFHT, Subaru, and by VLT
(\cite{vuds,vanzella10b}). Intrinsically faint galaxies ($L<<L^*$)
instead could be detected in LyC thanks to the large magnifications by
strong lensing, as proposed by
\cite{vanzella11} and already carried out by \cite{amorin14}. The
CLASH survey (\cite{clash}) and the Frontier Field Initiative by HST
will provide a large sample of highly magnified galaxies at high-z,
which will open a new window on the reionization process.

%__________________________________________________________________

\begin{acknowledgements}
We thank the anonymous referee for her/his useful suggestions and
comments that help us to improve this paper.
We acknowledge financial contribution from the agreement ASI-INAF I/009/10/0.
The LBT is an international collaboration among institutions in the
United States, Italy, and Germany. LBT Corporation partners are The
University of Arizona on behalf of the Arizona university system;
Istituto Nazionale di Astrofisica, Italy; LBT
Beteiligungsgesellschaft, Germany, representing the Max-Planck
Society, the Astrophysical Institute Potsdam, and Heidelberg
University; The Ohio State University; and The Research Corporation,
on behalf of The University of Notre Dame, University of Minnesota,
and University of Virginia.
This work is
supported by funding from the European Research Council Advanced Grant
ERC-2010-AdG-268107-EARLY and by INAF Grants PRIN 2010, PRIN 2012 and
PICS 2013. AC, OC, MT and VS acknowledge the grant MIUR PRIN 2010--2011.
DM gratefully acknowledges LAM hospitality during the initial phases of
the project. This work is based on data products made available at the
CESAM data center, Laboratoire d'Astrophysique de Marseille.
This work partly uses observations
obtained with MegaPrime/MegaCam, a joint project of CFHT and
CEA/DAPNIA, at the Canada-France-Hawaii Telescope (CFHT) which is
operated by the National Research Council (NRC) of Canada, the
Institut National des Sciences de l'Univers of the Centre National de
la Recherche Scientifique (CNRS) of France, and the University of
Hawaii. This work is based in part on data products produced at
TERAPIX and the Canadian Astronomy Data Centre as part of the
Canada-France-Hawaii Telescope Legacy Survey, a collaborative project
of NRC and CNRS.
Based on observations made with ESO Telescopes at the La Silla or
Paranal Observatories under programme ID 175.A-0839.
AG and EV warmly thank A. K. Inoue for providing the updated IGM
transmission used in this paper.
AF and JSD acknowledge
the contribution of the EC FP7 SPACE project ASTRODEEP (Ref.No:
312725).
\end{acknowledgements}

\Online

\begin{appendix}

\section{Other examples of contamination by foreground galaxies}

\begin{figure*}
\includegraphics[width=18cm,angle=0]{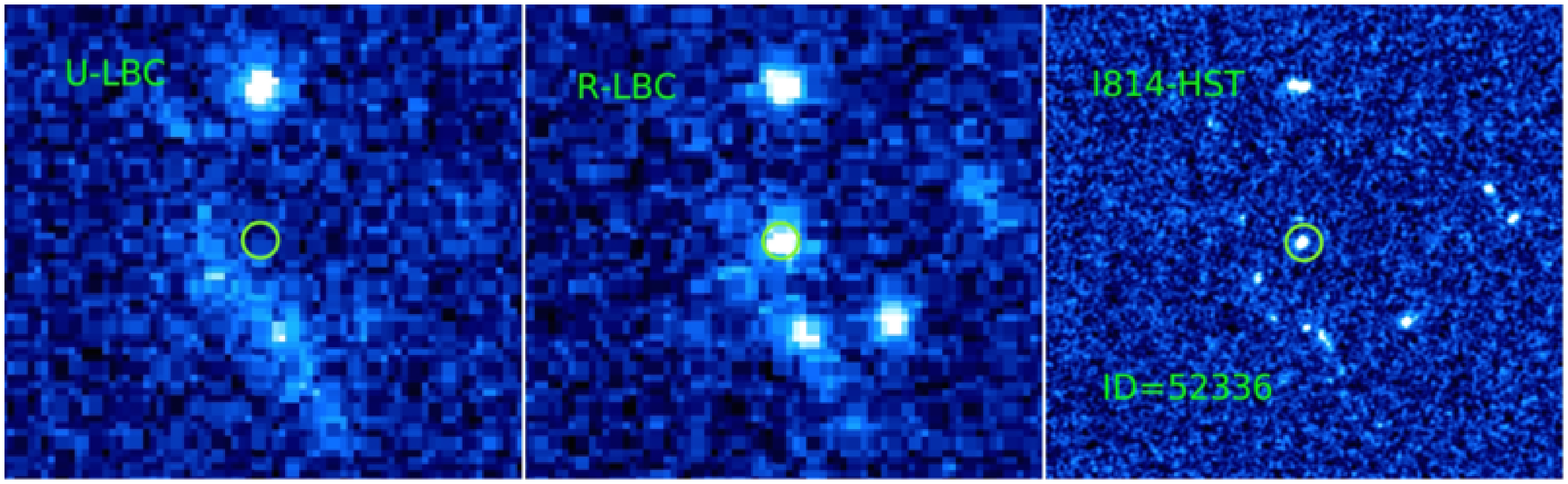}
\includegraphics[width=18cm,angle=0]{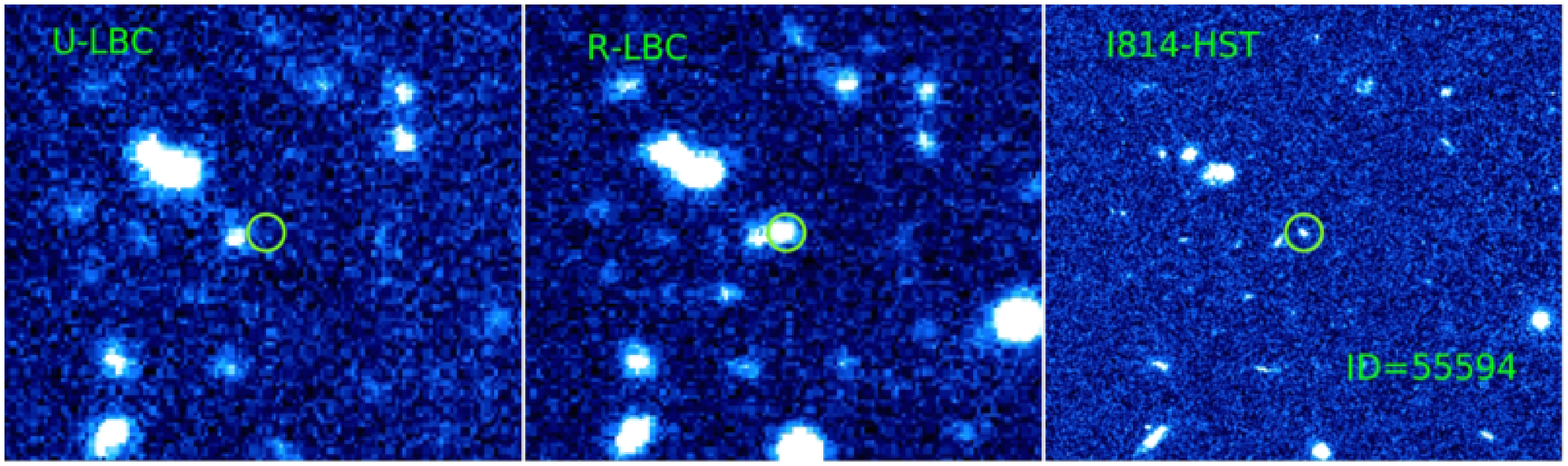}
\caption{The cutouts of sources ID=52336 and 55594 in the U band (left)
and in the
R band (central panels) by the LBC instrument. The right cutouts show the
same sky area observed by HST in the I814W filter. The positions of the
galaxies with spectroscopic redshifts $z\sim 3.3$ are indicated
by the green circles. The size of the cutouts is 12 arcsec for source
ID=52336 and 25 arcsec for ID=55594, respectively.
}
\label{obj2ima}
\end{figure*}

\begin{figure*}
\includegraphics[width=18cm,angle=0]{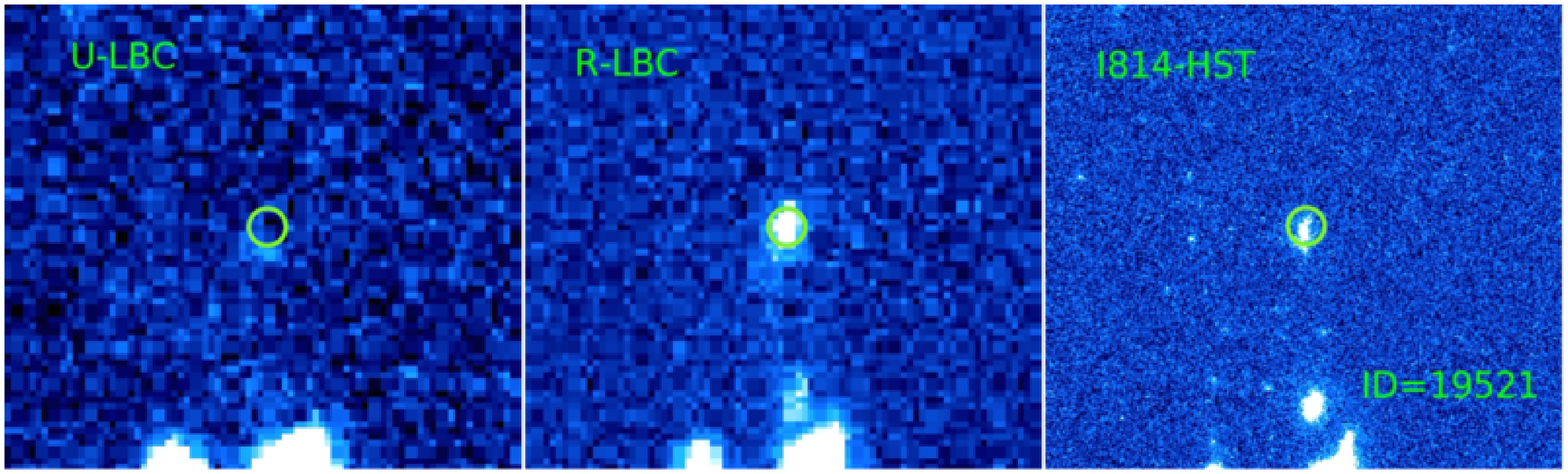}
\includegraphics[width=18cm,angle=0]{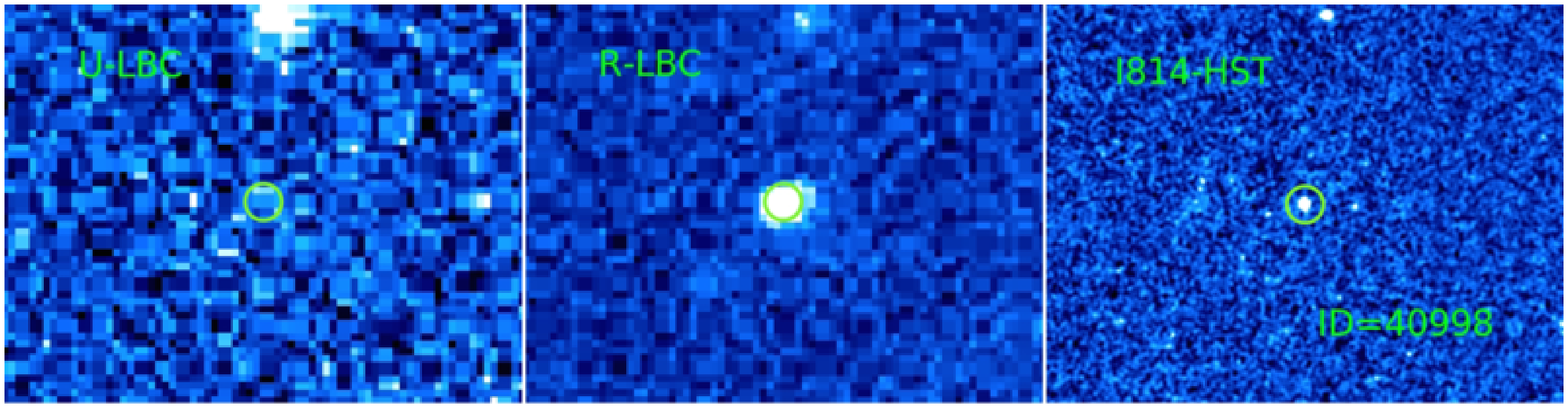}
\includegraphics[width=18cm,angle=0]{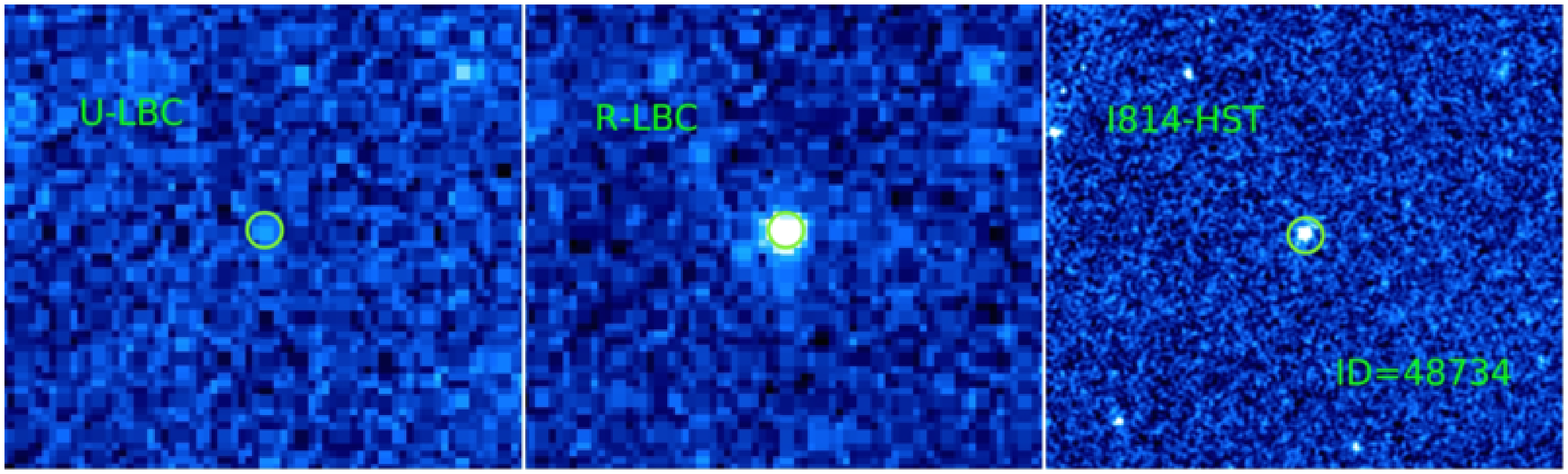}
\includegraphics[width=18cm,angle=0]{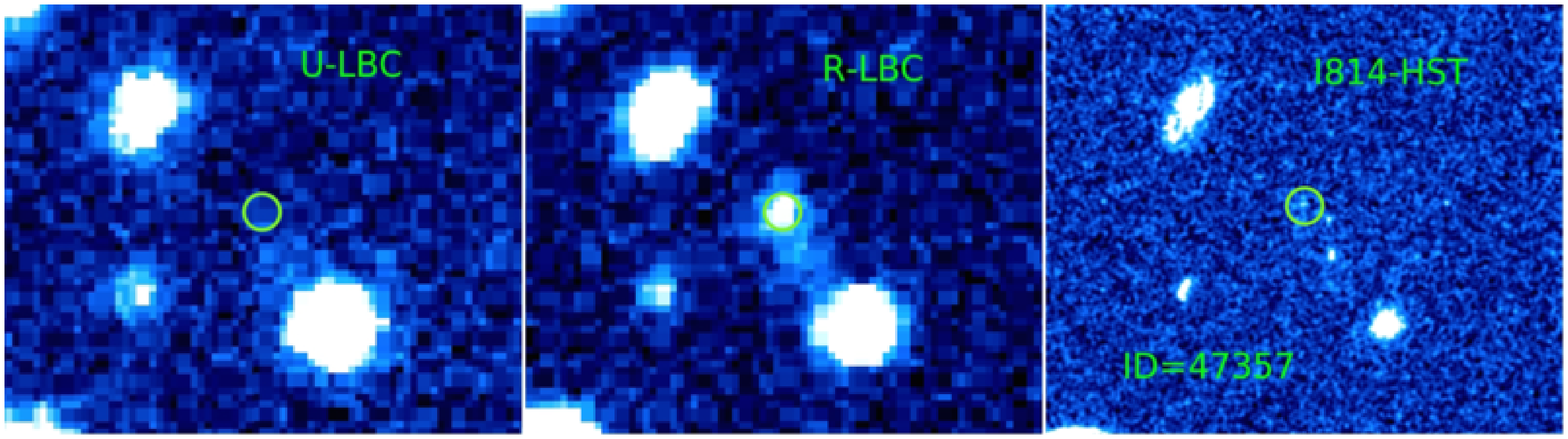}
\caption{The cutouts of sources ID= 19521,
40998, 48734, and 47357 in the U band (left) and in the
R band (central panels) by the LBC instrument. The right cutouts show the
same sky area observed by HST in the I814W filter. The positions of the
galaxies with spectroscopic redshifts $z\sim 3.3$ are indicated
by the green circles.
The size of the cutouts is 15, 10, 12, and 10 arcsec for source
ID=19521, 40998, 48734, and 47357, respectively.
}
\label{ima4obj}
\end{figure*}

We show here the cutouts in the U and R bands of LBC and in the I814W band
by HST of peculiar sources of the VUDS-LBC/COSMOS field. These sources,
in the last two subsections of Table \ref{table:gal}, have been detected
in LyC, but there are evidences of contamination
by foreground sources closely aligned with the $z\sim 3.3$ sources,
as mainly discussed in Section 4. In particular, Fig.\ref{obj2ima} shows two
objects with ``global'' $f^{rel}_{esc}\ge 100\%$ and Fig.\ref{ima4obj} shows the
example of four galaxies with ``global'' $f^{rel}_{esc}< 100\%$. In all the six
galaxies however the ``local'' escape fraction is in the range
$\sim 200-1000\%$, thus indicating contamination by foreground sources.

\end{appendix}

\end{document}